\documentclass[11pt, a4paper,onecolumn]{article}

\usepackage{graphicx}
\usepackage[T1]{fontenc}
\usepackage{geometry}
\usepackage{amsmath}
\usepackage[utf8]{inputenc}
\usepackage{fontspec}
\usepackage{pdflscape}
\usepackage{xcolor}
\usepackage{amssymb}
\usepackage[font=footnotesize,labelfont=bf]{caption}
\usepackage{graphicx}
\usepackage[round,sort&compress]{natbib}
\usepackage{tabularx}
\usepackage{url}

\usepackage[colorinlistoftodos]{todonotes}

\usepackage{booktabs}

\begin{document}

\onecolumn
\vspace*{3cm}
{\LARGE Spatial accessibility to food banks hinders food parcel uptake in England and Wales, particularly in rural areas}\\

Laura Sheppard$^{1}$, Carmen Cabrera$^{2}$, Daphne Badounas$^{1,5}$, Bonnie Boyana Buyuklieva$^{3}$, Sukankana Chakraborty$^{1}$, Huanfa Chen$^{1}$, Sarah Wise$^{1}$,Howard Wong$^{1,5}$, Rachael Jones$^{4}$, Neave O'Clery$^{1}$.

\begin{footnotesize}
1. Centre for Advanced Spatial Analysis, University College London \newline
2. Geographic Data Science Lab, Department of Geography and Planning, University of Liverpool \newline
3. Department for Information Studies, University College London. \newline
4. Trussell \newline
5. Transport for London (TfL) 
\end{footnotesize}

\vspace{0.3cm}
\noindent\rule{\linewidth}{0.35pt}
\vspace{0.3cm}

\setlength{\leftskip}{2.5cm}

A B S T R A C T

\setlength{\leftskip}{0cm}

\vspace{0.3cm}
\noindent\rule{\linewidth}{0.35pt}
\vspace{0.3cm}

\setlength{\leftskip}{2.5cm}
Food bank use in the UK has soared in recent years. The combination of a global pandemic, over-stretched and underfunded public services, and a cost-of-living crisis has meant that millions of people cannot afford basic essentials such as food, heating, housing, and baby supplies.
Food bank use is driven by a complex range of factors, including poverty, health emergencies, income shocks, delays to universal credit payments, housing issues, and homelessness.
In this study we identify an urban-rural divide in spatial accessibility to food banks. In cities, food banks tend to be highly accessible by public transport to deprived populations but, on average, have shorter opening hours. In rural areas, however, despite generally longer opening hours, food banks are typically not highly accessible except for the most deprived residents. 
This matters. We find that spatial accessibility to a Trussell food bank centre is a key predictor of food parcel uptake, with a significantly stronger relationship than factors emphasised in the literature such as disability and Universal Credit. Importantly, this relationship is markedly stronger for rural populations, suggesting an unmet need in deprived rural areas far from food banks. 
Our work has important implications for food bank policy, suggesting a need for improved public transport in rural areas, and optimising current food bank locations and delivery models. 
\\
\\
{\it JEL Codes}: XXX. \\
{\it Keywords}: Food banks, Food poverty, Inequalities, Accessibility, Trussell \\

\setlength{\leftskip}{0cm}
\section*{Acknowledgments and funding}

NOC, LS and RJ would like to acknowledge funding from the UCL 2022-23 Grand Challenges Special Initiative on Food Security. 

\clearpage

\section{\bf{Introduction}}

By mid 2024, it was estimated that 7.2 million adults (13.6\% of the UK population) were food insecure \citep{Food-Foundation-2024}. Food insecurity relates to reduced food quality, variety, and choice, and often results in disrupted eating patterns, skipped meals, or reduced overall food intake \citep{Francis-Devine-2024}. Food insecurity and a reliance on emergency food parcels can impact an individual's capacity to work and study, and contributes to malnutrition, and poor mental and physical health outcomes \citep{Safayet-2024}. The use of food banks has increased significantly in the UK since 2010, in part due to austerity, the 2008 global financial crisis, and changes to social policy and welfare reform \citep{Lambie-Mumford-2019, Loopstra-2019}. However, there has been an even higher growth in food bank use since 2020, driven by the global Covid-19 pandemic, rising inflation rates, and a persistent cost-of-living crisis \citep{Hunger-in-the-UK}.    

Food banks are charitable organisations that collect and distribute food to people in need in their local communities. The UK is home to over 2,500 food bank distribution centres which are typically run as single organisations or in local clusters. Most UK food banks are members of one of the two national advocacy organisations: Trussell (previously known as The Trussell Trust) and the Independent Food Aid Network (IFAN) \citep{Zaidi-2025}. Covering approximately half of UK food banks, Trussell supports a community of over 400 food banks in a range of ways including operational and strategic support \citep{Bramley-2021}. Trussell advocates for ending the need for food banks at a national level, while aiming to meet current need. The size of the Trussell network has been relatively stable since the pandemic, but the number of parcels given out has risen sharply. In 2010/11, Trussell food banks distributed 61,000 emergency food parcels, but by 2023/24 this had risen to over 3.1 million food parcels. This represents an astronomical increase of 94\% over just five years \citep{Trussell-End-of-year-stats}. 

To receive a food parcel from a Trussell centre, individuals or families must have a referral voucher. Vouchers are issued by someone like a doctor, social worker, or teacher, or, alternatively, a social advice agency like Citizens Advice. Food banks outside of the Trussell network, however, do not necessarily require a referral. The typical contents of an emergency food parcel (one per person) include a combination of long-life food items such as milk, tinned vegetables, pulses, and fish/meat, as well as non-food items, which are discretionary and do vary between food banks \citep{Lambie-Mumford-2019}.

The reasons that put people at risk of hunger and hardship and may need to use food banks are interrelated. The 2023 \textit{Hunger in the UK} report suggests that people with disabilities, families with children, people that are unemployed and groups who experience structural inequalities such as LGBTQ+, asylum seekers, or ethnic minorities, are most at risk. In particular, delays or changes in income from benefits like universal credit are key drivers of food bank use \citep{Prayogo-2018}. \citet{Sosenko-2022} found that a £1 a week increase in the universal credit standard allowance is associated with a 2.6\% annual decrease in food bank use. \citet{Reeves-Loopstra-2021} suggest that the relationship between universal credit use and food parcel distribution is stronger in areas where food banks are active, suggesting that there is likely significant food insecurity in places without active food banks. Similarly, with over two thirds of UK households registered as having at least one person disabled \citep[p.~11]{Bramley-2021}, disability and poor health are key drivers of food bank use. Food insecurity, poor health, and food bank use can be cyclical, with poor nutrition driving increased ill health and future food bank use \citep[p.1204]{Oldroyd-2022}. 

Our focus, living close to key services, like health, food, and retail, is important for individuals to met their basic needs. Various studies have shown that accessibility to amenities and services - such as banks \citep{Small-2021, Hegerty-2016}, food shops \citep{Walker-2010}, green spaces \citep{Mears-2019}, doctors surgeries and pharmacies \citep{Todd-2015} - is impacted by transport options, opening hours, and location. Here we investigate drivers of spatial accessibility to food banks.

Previous work has focus primarily on cities, showing that low income neighbourhoods tend to have better geographic accessibility to food banks. In Berlin, the geographical distribution of food pantries is largely identical to the distribution of welfare recipients \citep[p.~95]{Simmet-2017}, with over a third of all welfare recipients within walking distance of a food pantry. In Toronto, new food bank distribution centres were found to increase the share of low-income residents within twenty minutes on public transport to a food bank \citep{Allen-Farber-2021}, with better accessibility in areas with a higher share of people receiving state welfare. These findings are generally consistent with research looking at accessibility to a variety of other services and amenities like primary schools and public nurseries and outdoor play areas \citep{Macdonald-2016, Macintyre-2008}. 

However, a focus on cities in previous research may mask important urban-rural divides in spatial accessibility to services. In other words, to what extent is there good access to services for low-income groups in rural areas. \citet{Smith-2010} suggest that urban neighbourhoods have better access to food shops than rural locations, particularly in highly deprived areas. This is echoed in access to other services like general practice surgeries. \citet{Todd-2015} found that while 98\% of people living in the most deprived areas in England live within a 20 minute walk of a doctor’s surgery, this is the case for only 19\% of people in rural areas. Here we seek to better understand the role of spatial accessibility (measured as either public transport travel time or distance) in food parcel uptake, and the extent to which this varies in urban and rural settings. We focus on in-person access to food bank distribution centres, and do not include delivery-only centres. We also consider food bank opening times as a second dimension of accessibility and how this varies between urban and rural areas.

Trussell data suggests that in urban areas, as we would expect, food parcels and food bank centres are more likely to be concentrated in deprived areas. There is a significant urban-rural divide, however, with the most to moderately deprived urban areas having high accessibility (around 30 minutes), but this is only true for the most deprived rural places. Deploying a multivariate econometric model at lower super output area (LSOA) level, we find that key variables from the literature (such as Universal Credit claimants, socially rented homes, and disability) are robustly associated with higher food parcel uptake. However, we find that accessibility in terms of travel time and distance is a much stronger determinant of food parcel uptake relative to these factors. Rural inhabitants are more sensitive to food bank accessibility relative to urban dwellers, suggesting that overall access to rural food banks is worse. 

Our work has important takeaways. Fundamentally, we show that spatial accessibility is a key but under-appreciated determinant of food bank use. In essence, there are people going without the essentials who need to turn to a food bank but are unable to travel to one because of time, distance or public transport availability, particularly in rural settings. This has immediate implications for both the positioning and opening times of food bank centres, and the provision of public transport or other transport options. 

\section{\bf{Data and Methods}}

\subsection{Data from Trussell}

England and Wales are the chosen study area due to data coverage and availability. The food parcel data as well as the addresses and postcodes of all Trussell centres and information on their operational service models was provided by Trussell's Strategic Intelligence Team. The Trussell food parcel data for England and Wales was collected during the 2021/22 financial year and is at the scale of the Lower Super Output Area (LSOA). An LSOA is a small geographical unit that contains 400-1200 households and 1000-3000 residents. In total there are 35,726 LSOAs in England and Wales \citep{Statistical-geographies}. We use LSOA centroids as the home location of the person receiving the food parcel.

Our dependent variable is food parcels per one thousand residents for each LSOA. People can receive more than one food parcel a year, and hence some people will be counted more than once in these statistics. 

The data processing steps are outlined in more detail in the SI, and they include removing centres that were closed, delivery only, did not directly distribute food like offices and warehouses. A total of 1,187 (from 1,300) Trussell distribution centres remain as part of our analysis.

\subsection{Census data and other sources of data}

We used Functional Urban Areas \citep{Dijkstra-2019} to classify areas as either urban or rural. Approximately two thirds of LSOAs fall within Functional Urban Areas and were classified as urban, and the other one third were classified as rural.

From the academic literature and recent Trussell reports, we identified the main known determinants of UK food bank use \citep{Hunger-in-the-UK, Bramley-2021, Francis-Devine-2024, Reeves-Loopstra-2021, Loopstra-2019, MacLeod-2018, Wainwright-2018, Prayogo-2018, Sosenko-2022}. These factors include: disability, poor health status, high deprivation, low income, universal credit claimants, part time work and economic inactivity, providing unpaid care, large household size, households with children, socially renting, and lone parent households. The majority of these variables were derived from the 2021 Census. Data for Universal Credit claimants was accessed through the Department for Work and Pensions's \textit{Stat-Xplore} site \citep{DWP-2024}. A full list of variables is included in Tables \ref{tab:census_variables} and \ref{tab:additional_variables} in the SI. 

Some of the variables such as universal credit claimants and economic inactivity are likely to overlap and so we chose a regression method (see below) that explicitly addresses this overlap. Figure \ref{fig:Fig_corr} of the SI presents pairwise correlations between the chosen variables. Our dependent variable (parcels per 1000 residents) is positively correlated with the share of Universal Credit claimants, lone parent families, bad health, disability, deprivation and social renting, and having no qualifications. Overall, we observe blocks of high correlation between variables related to health and education; health, disability and education; and Universal Credit claimants, lone parents, no car ownership and social renting. Conversely, some variables are negatively correlated, such as having a high share of highly qualified residents (level 4 qualifications) with bad health and disability indicators.

Finally, the Index for Multiple Deprivation (IMD) 2019 was used to group LSOAs by deprivation deciles in our visualisations. We note that aggregated measures such as the IMD can hide some pockets of deprivation in (more likely rural) larger size LSOAs. 

\subsection{Accessibility modelling}

We aim to measure spatial accessibility to food bank distribution centres. This is 'potential' accessibility in the sense that it captures the 'cost' in time and distance to travel to a food bank, but does not try to directly estimate actual journeys taken. For each LSOA, we compute the distance or time to the closest food bank via different modes of transport (walking, driving, and public transport). Taking multiple measures of accessibility into account in our regression models allows us to examine the potential differences in impact they may have on food bank use.  

First, we compute the minimum (euclidean) distance from the centroid of each LSOA to the closest food bank centre (in kilometres). Second, we compute walking and driving times from the centroid of each LSOA to the closest food bank centre using the the R package r5r \citep{r5r}. Third, we compute the minimum public transport travel time from the centroid of each LSOA to the closest Trussell food bank centre. To do this, we use public transport travel times from the R package r5r \citep{r5r} in combination with General Transit Feed Specification (GTFS) timetable data (in minutes).

We note that, as travel time calculations use the LSOA centroid, the differences in size of the urban versus rural LSOAs may affect these calculations as urban LSOAs tend to have a much smaller area than rural ones.

\subsection{Regression modelling}

We use multi-variate regression modelling to understand the relationship between accessibility, other variables in the literature and food bank use in England and Wales. Our model specification is: 
$$
Y_i=\beta_0 + \beta_1X_1i + \beta_2X_2i + ... + \epsilon
$$
for LSOA numbered $i$, outcome variable $Y_i$ food parcels per 1000 residents, intercept $\beta_0$, explanatory variables $X_i$, and constant $\epsilon$. All variables are standardised. 

In order to mitigate for the high level of inter-correlation between our explanatory variables, we adopt a step-wise regression approach \citep{Marascuilo-1988, Walter-2009}. This approach uses statistical significance of each variable (specifically the variance inflation factor, or VIF, was used to detect multi-collinearity) to remove interrelated and redundant variables that do not add to the overall predictive power of the model. We present the full step-wise analysis in the SI (Tables \ref{stepwise_regression_1}, \ref{stepwise_regression_2}, and \ref{stepwise_regression_3}), while Model 1 of Table \ref{standardised_regression_results} shows the final result of this process.

\newpage

\section{\bf{Results}}

We investigate the distribution of, and spatial accessibility to, food banks in England and Wales. We then investigate the importance of spatial accessibility in the uptake of food parcels relative to a range of existing factors in the literature.

\subsection{\bf{Geography of food banks and food parcels}}
 
\begin{figure}[!t]
    \centering
    \includegraphics[width = 14.5cm]{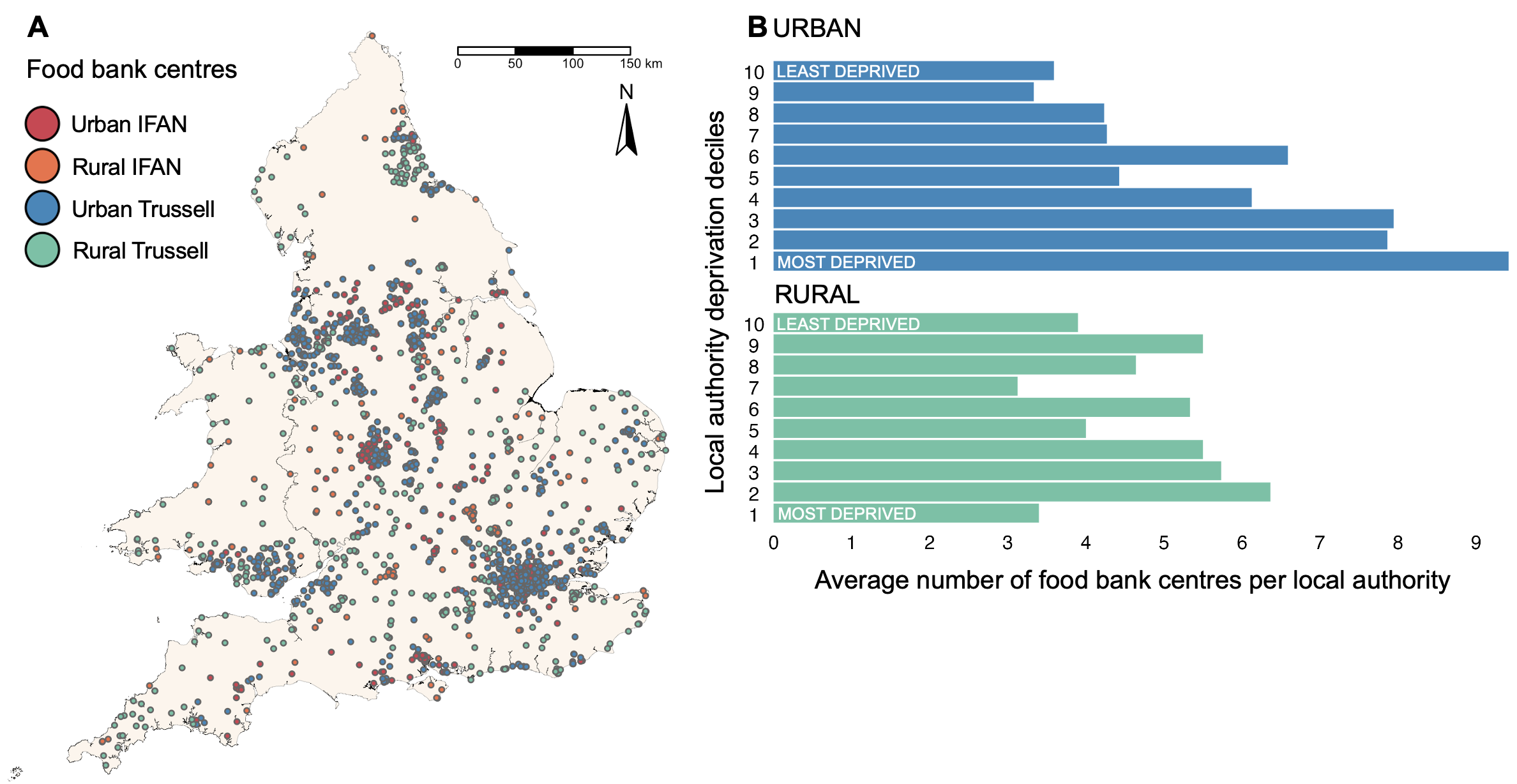}
    \caption{[A] We observe a dense spatial clustering of food banks in major cities and urban areas, with a similar concentration of Trussell and IFAN food banks across urban/rural areas. [B] Food bank centres tend to be located in more deprived areas in urban areas, however this is not true for rural settings.}
    \label{fig:Fig1}
\end{figure}

In Figure \ref{fig:Fig1}A, we observe a dense spatial clustering of food banks in major cities and urban areas, with a significant presence in some rural areas like Cheshire, County Durham and Cornwall. Overall, the mean number of food banks in urban local authorities is 3.8, while its 1.9 in rural local authorities. Both Trussell and IFAN food banks have a similar concentration across urban and rural areas. 

While it is not surprising that food banks would locate in densely populated urban centres, to what extent are they serving the most deprived areas? To investigate this, we harness the 2019 Index for Multiple Deprivation (IMD) to assign local authorities to a social deprivation decile from 1 (highest 10\%) to 10 (lowest 10\%). 
Figure \ref{fig:Fig1}B highlights a distinct contrast between urban and rural local authorities. Specifically, while food bank centres are disproportionately located in deprived local authorities in urban areas, this is not true for rural settings. In other words, the more deprived an urban area is, the more likely it is to have a food bank centre in it. But this does not hold true for rural areas which show no easily discernable pattern. This preliminary analysis suggests that, in rural areas, more deprived populations are more likely not to live in close proximity to a food bank. 

\begin{figure}[!ht]
    \centering
    \includegraphics[width = 14cm]{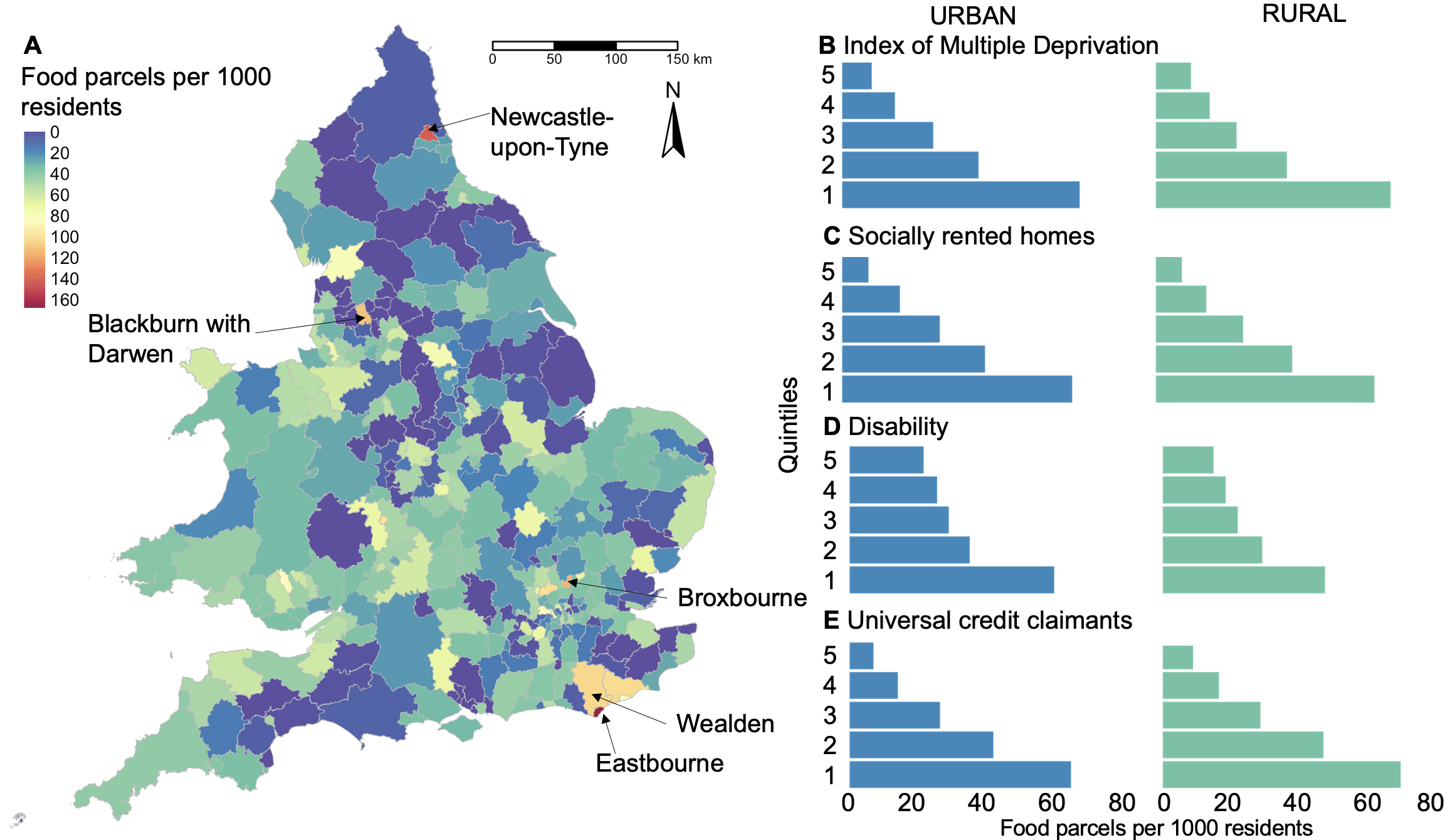}
    \caption{[A] We observe scattered hotspots of food parcel uptake from Newcastle in the North to Eastbourne in the South. [B-E] For both urban and rural settings, food parcels are more likely to be distributed in areas with higher levels of overall deprivation, socially rented homes and universal credit claimants. Food parcel uptake is also concentrated areas very high levels of disability.}
    \label{fig:Fig2}
\end{figure}

If food banks are more likely to be located in deprived areas in cities, do we see the same pattern in terms of food parcel uptake? 
Figure \ref{fig:Fig2}A shows that food parcel uptake is also spatially clustered, with high uptake in areas such as Eastbourne, Newcastle-upon-Tyne, Broxbourne, and Blackburn with Darwen. These areas are scattered across the country\footnote{The spatial distribution of the food parcels is not random: the Moran's I test for spatial auto-correlation at the local authority level was 0.132 (Moran's I ranges between -1 and 1), showing that there is a slight positive spatial auto-correlation for food parcel distribution. This was statistically significant to 95\%}, and span across the deprivation ranks. 

To investigate the extent to which food parcel uptake is higher in deprived areas, we segment LSOAs into five quintiles (quintile 1 has the most deprived 20\% of LSOAs, whereas quintile 5 has the least).
Overall, across both urban and rural areas, we observe that LSOAs with higher overall deprivation also receive more food parcels (Figure \ref{fig:Fig2}B). This holds true for socially rented homes and universal credit claimants. 
However, this relationship is more complex for disability, with a weaker overall increase in food parcels as disability increases. 

\subsection{\bf{Accessibility and opening hours}}

\begin{figure}[!ht]
    \centering
    \includegraphics[width = 14cm]{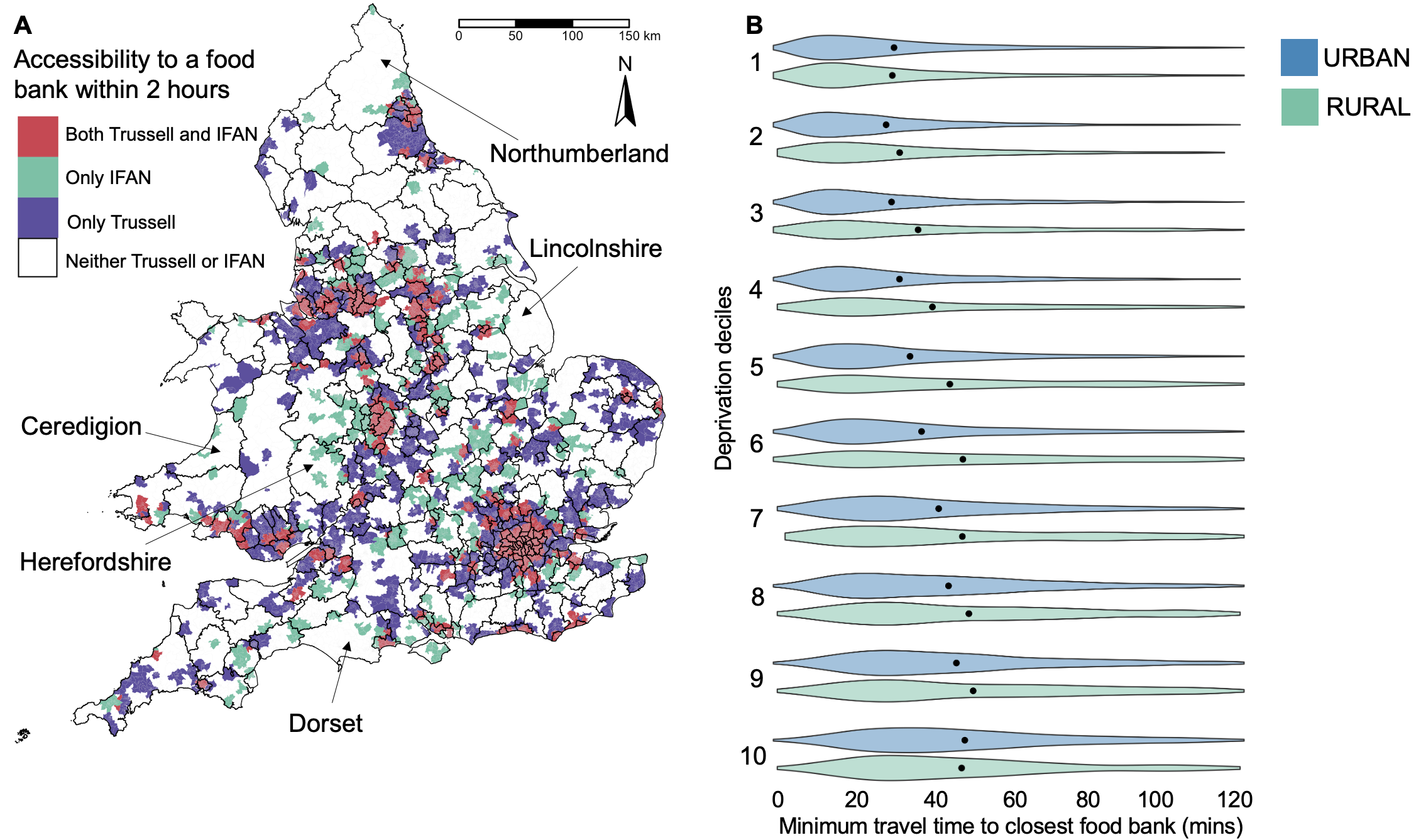}
    \caption{[A] We observe significant holes in which neither a Trussell or IFAN food bank is accessible within two hours on public transport. [B] We find that the most deprived areas have the highest accessibility to their closest food bank (30 minutes in decile 1) for both urban and rural places. However, the gap between accessibility for urban and rural areas increases as deprivation deciles increase and is largest in the moderately deprived areas.}
    \label{fig:Fig3}
\end{figure}

Building on this, we next investigate the extent to which people have access to a food bank centre within two hours. Figure \ref{fig:Fig3}A shows that large areas (white space on the map) have no public transport to a Trussell or IFAN food bank within two hours, including in large parts of Dorset, Lincolnshire, and Northumberland which are all predominately rural local authorities. We note that some of these local authorities may be served by delivery-only food bank centres, which we do not capture here.
In general, we see good spatial accessibility to both Trussell and IFAN food banks in larger cities such as London, Liverpool, Manchester, and Birmingham.

We expect that rural LSOAs experience a lower overall accessibility to food bank centres. Here we test that hypothesis.
Figure \ref{fig:Fig3}B shows the distribution (and mean) of minimum public transport travel times to an area's closest food bank centre (Trussell or IFAN) by deprivation decile (for median travel time, see Figure \ref{fig:Fig3_appendix} in the SI). 
In urban areas, we observe a low mean time between 28 to 34 minutes for the first five deciles (higher deprivation) before slowly increasing to nearly 50 minutes in the least deprived decile. 
For the most deprived rural areas (decile 1), the mean time is the same as urban areas at 30 minutes. However, the gap between rural and urban mean travel times quickly increases as the deprivation deciles increase. For example, by decile 5 (moderate deprivation), there is a difference of 10 minutes (44 minutes for rural, and 34 minutes for urban). From decile six onwards there is a plateau and the mean hovers at around 50 minutes.
Pointing to a concerning gap in access, overall we observe a significant urban-rural divide in that most deprived urban areas (ranging from moderately to very deprived) have high accessibility to their closest centre, but only the most deprived rural ones have the same. 

\begin{figure}[!th]
    \centering
    \includegraphics[width = 14cm]{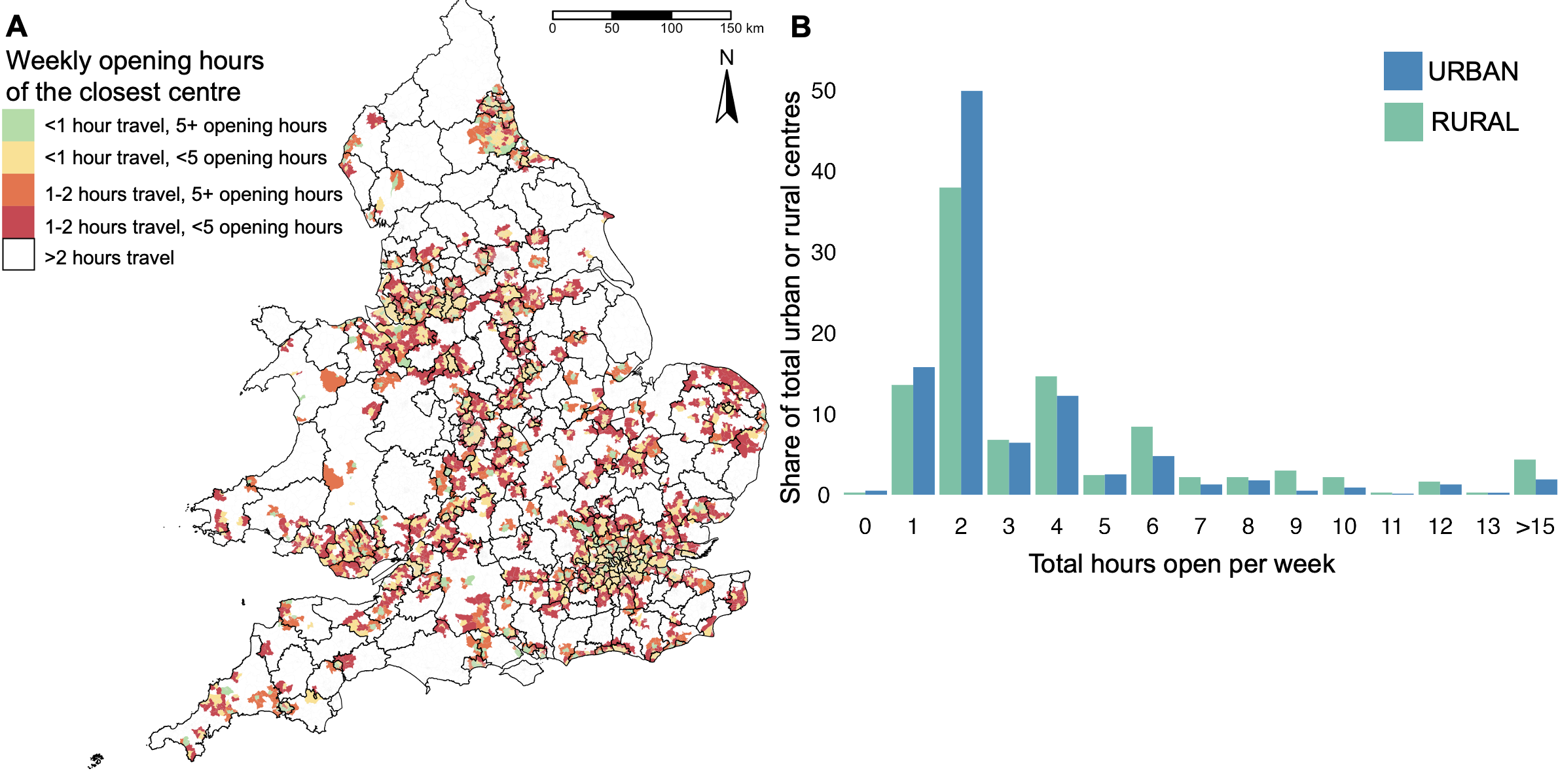}
    \caption{[A] Urban areas are more likely to have access to a Trussell food bank within two hours public transit travel time - but not necessarily longer opening hours (per week). [B] Nearly 50\% of urban food bank centres are open for only two hours a week, whereas rural centres are more likely to have longer opening hours.}
    \label{fig:Fig4}
\end{figure}

Focusing on Trussell food bank centres, we consider a second dimension of accessibility - food bank opening times. The opening hours for Trussell food banks, in 2019, tended to be concentrated between the hours of 10 am and 2 pm, with only 13.5\% of areas with a food bank centre open in the evening \citep[p.~4]{Loopstra-2019}. Our data suggests that rural centres are open 4.3 hours and 1.8 days a week, compared to urban centres at 3.4 hours and 1.5 days a week.
As shown in Figure \ref{fig:Fig4}A, only 12\% of LSOAs have good access in terms of travel time less 1 hour and opening hours > 5 hours. 
In general, urban areas tend to have good travel time but generally shorter centre opening hours, whereas rural areas have longer travel times but their centres are, on average, open for longer.
There are observable hotspots of local authorities with longer total opening hours. For example, the combined weekly opening hours of all food bank centres in Dacorum (which has a food bank housed in Dacorum Borough Council buildings), County Durham, Salford, Cheshire West and Chester, and Cornwall, exceeds 100 hours. 
Figure \ref{fig:Fig4}B supports these observations, showing that nearly 50\% of urban Trussell centres are open for only two hours a week, whereas rural centres are more likely to have longer opening hours with some open for over 15 hours per week. 

\subsection{\bf{Determinants of food bank use}}

We have seen above that the concentration of food banks scales heavily with deprivation in urban areas but less so in rural areas, and food parcel uptake increases with deprivation decile. Here we deploy econometric methods to investigate the full set of correlates of food parcel uptake, with a particular focus on accessibility. 

Regression model 1 shows that, consistent with the literature, a large number of variables are positive and statistically significant correlates of food parcel use: disability, share of universal credit claimants, socially rented homes, lone parents, level one qualifications, and part time workers. For the highest positive coefficient value, a one standard deviation (sd) increase in households with one or more persons who are disabled is associated with a 0.20 sd increase in food parcels. Translating this to interpretable figures, a 15.9\% increase in households with one or more people who are disabled is associated with an increase in food parcel use of 11.8 parcels per one thousand residents. 
Similarly, we find that a one sd increase in universal credit claimants per residents is associated with a 0.17 sd increase in food parcels - or a 2.2\% increase in claimants is associated with an increase of 10 parcels by one thousand residents.
The size and significance of the coefficients are very similar for urban LSOAs (model U1 of Table \ref{regression_urban_standardised}), while for rural LSOAs (model R1 of Table \ref{regression_rural_standardised}) we have a larger coefficient for lone parents, but lower for disability.
Overall, the total variance (r2) explained by Model 1 (the base model) is 0.179 (0.182 for urban areas and 0.169 for rural areas).

\begin{table}[!t] \centering 
  \caption{Multiple regression models, with dependent variable  parcels per 1000 residents, with standardised explanatory variables and accessibility measures.} 
  \label{standardised_regression_results}
  \tiny
\begin{tabular}{lccccccc} 
 & \multicolumn{7}{c}{\textit{Dependent variable: Parcels per 1000 residents}} \\ 
\cline{2-8} 
\\[-1.8ex] & 1 & 2 & 3 & 4 & 5 & 6 & 7 \\ 
\hline \\[-1.8ex] 
 Universal Credit claimants (\% of people) & 0.175$^{***}$ & 0.232$^{***}$ & 0.162$^{***}$ & 0.179$^{***}$ & 0.164$^{***}$ & 0.161$^{***}$ & 0.164$^{***}$ \\ 
  & (0.007) & (0.006) & (0.007) & (0.007) & (0.006) & (0.007) & (0.007) \\ 
  
 Single family HHs with a lone parent (\%) & 0.091$^{***}$ & 0.094$^{***}$ & 0.089$^{***}$ & 0.090$^{***}$ & 0.036$^{***}$ & 0.029$^{***}$ & 0.075$^{***}$ \\ 
  & (0.009) & (0.009) & (0.009) & (0.009) & (0.009) & (0.009) & (0.010) \\ 
  
 Socially rented HHs (\%) & 0.139$^{***}$ & 0.136$^{***}$ & 0.139$^{***}$ & 0.140$^{***}$ & 0.158$^{***}$ & 0.168$^{***}$ & 0.145$^{***}$ \\ 
  & (0.008) & (0.009) & (0.008) & (0.008) & (0.008) & (0.008) & (0.009) \\ 
  
 HHs with 1+ person with a disability (\%) & 0.185$^{***}$ & 0.150$^{***}$  & 0.163$^{***}$ & 0.186$^{***}$ & 0.164$^{***}$ & 0.166$^{***}$ & 0.191$^{***}$ \\ 
  & (0.010) & (0.010) & (0.010) & (0.010) & (0.010) & (0.011) & (0.011) \\ 
  
 Economically inactivity (\% of people) & $-$0.063$^{***}$ & $-$0.033$^{***}$ & $-$0.059$^{***}$ & $-$0.063$^{***}$ & $-$0.069$^{***}$ & $-$0.066$^{***}$ & $-$0.073$^{***}$ \\ 
  & (0.009) & (0.008) & (0.009) & (0.009) & (0.008) & (0.009) & (0.009) \\ 
  
 Part time workers (\% of people) & 0.060$^{***}$ & 0.001 & 0.063$^{***}$ & 0.060$^{***}$ & 0.111$^{***}$ & 0.106$^{***}$ & 0.086$^{***}$ \\ 
  & (0.007) & (0.008) & (0.007) & (0.007) & (0.007) & (0.008) & (0.008) \\ 
  
 Level 1 qualification (\% of people) & 0.085$^{***}$ & 0.094$^{***}$ & 0.084$^{***}$ & 0.084$^{***}$ & 0.102$^{***}$ & 0.127$^{***}$ & 0.112$^{***}$ \\ 
  & (0.008) & (0.009) & (0.008) & (0.008) & (0.008) & (0.009) & (0.009) \\ 
  
 Level 2 qualification (\% of people) & $-$0.044$^{***}$ & $-$0.051$^{***}$ & $-$0.044$^{***}$ & $-$0.044$^{***}$ & $-$0.023$^{***}$ & $-$0.020$^{**}$ & $-$0.046$^{***}$ \\ 
  & (0.008) & (0.008) & (0.008) & (0.008) & (0.008) & (0.008) & (0.008) \\ 
  
 Apprenticeships (\% of people) & $-$0.058$^{***}$ & $-$0.013$^{**}$ & $-$0.050$^{***}$ & $-$0.058$^{***}$ & $-$0.031$^{***}$ & $-$0.036$^{***}$ & $-$0.060$^{***}$ \\ 
  & (0.008) & (0.008) & (0.008) & (0.008) & (0.007) & (0.008) & (0.008) \\ 
  
 Level 3 qualification (\% of people) & $-$0.007 & $-$0.011$^{*}$ & $-$0.010$^{**}$ & $-$0.007 & $-$0.021$^{***}$ & $-$0.023$^{***}$ & $-$0.017$^{***}$ \\ 
  & (0.005) & (0.005) & (0.005) & (0.005) & (0.005) & (0.005) & (0.005) \\ 
  
 20 + hours of unpaid care (\% of people) & $-$0.096$^{***}$ & $-$0.070$^{***}$ & $-$0.084$^{***}$ & $-$0.097$^{***}$ & $-$0.100$^{***}$ & $-$0.120$^{***}$ & $-$0.100$^{***}$ \\ 
  & (0.009) & (0.009) & (0.009) & (0.009) & (0.009) & (0.009) & (0.010) \\ 
  
 Larger HHs (\% of people) & $-$0.018$^{**}$ & 0.031$^{***}$ & $-$0.016$^{**}$ & $-$0.019$^{**}$ & $-$0.038$^{***}$ & $-$0.037$^{***}$ & $-$0.027$^{***}$ \\ 
  & (0.008) & (0.007) & (0.007) & (0.008) & (0.007) & (0.008) & (0.008) \\ 
  
 Density of the LSOA & $-$0.047$^{***}$ & 0.005 & $-$0.041$^{***}$ & $-$0.048$^{***}$ & $-$0.069$^{***}$ & $-$0.090$^{***}$ & $-$0.049$^{***}$  \\ 
  & (0.006) & (0.007) & (0.006) & (0.006) & (0.006) & (0.006) & (0.007) \\ 
  
 Presence of Trussell centre &  &  & 0.158$^{***}$ &  & 0.130$^{***}$ & 0.115$^{***}$ & 0.152$^{***}$ \\ 
  &  &  & (0.005) &  & (0.005) & (0.005) & (0.005) \\ 
 Presence of IFAN centre &  &  &  & $-$0.034$^{***}$ & $-$0.023$^{***}$ & $-$0.020$^{***}$ & $-$0.032$^{***}$ \\ 
  &  &  &  & (0.005) & (0.005) & (0.005) & (0.005) \\ 
 Minimum distance &  &  &  &  & $-$0.234$^{***}$ \\ 
 &  &  &  &  & (0.005) & \\ 
 Minimum public transport travel time &  &  &  &  &  & $-$0.250$^{***}$\\ 
 &  &  &  &  &  & (0.005) \\
 Weekly opening hours &  &  &  &  &  &  & 0.025$^{***}$ \\ 
 &  &  &  &  &  &  & (0.005) \\
 Constant & 0.000 & 0.417$^{***}$ & $-$0.000 & $-$0.000 & $-$0.000 & 0.039$^{***}$ & 0.030$^{***}$ \\ 
  & (0.005) & (0.126) & (0.005) & (0.005) & (0.005) & (0.005) & (0.005) \\ 
\hline \\[-1.8ex] 
Fixed effects & No & Yes & No & No & No & No & No \\
R$^{2}$ & 0.176 & 0.350 & 0.201 & 0.177 & 0.248 & 0.259 & 0.210 \\ 
Adjusted R$^{2}$ & 0.176 & 0.344 & 0.200 & 0.177 & 0.248 & 0.258 & 0.209 \\ 
Residual Std. Error & 0.908 & 0.810 & 0.893 & 0.905 & 0.867 & 0.883 & 0.913 \\ 
DF & 35780 & 35658 & 35657 & 35657 & 35655 & 32562 & 31699 \\ 
\hline \\[-1.8ex] 
\textit{Note:}  & \multicolumn{2}{r}{$^{*}$p$<$0.1; $^{**}$p$<$0.05; $^{***}$p$<$0.01} \\ 
\end{tabular} 
\end{table}

\begin{table}[!t] \centering 
  \caption{Multiple regression models for urban LSOAs, with dependent variable  parcels per 1000 residents, with standardised explanatory variables and accessibility measures.} 
  \label{regression_urban_standardised} 
    \tiny
\begin{tabular}{lccccccc} 
 & \multicolumn{7}{c}{\textit{Dependent variable: Parcels per 1000 residents}} \\ 
\cline{2-8} 
\\[-1.8ex] & U1 & U2 & U3 & U4 & U5 & U6 & U7 \\ 
\hline \\[-1.8ex] 
  Universal Credit claimants (\% of people) & 0.173$^{***}$ & 0.234$^{***}$ & 0.162$^{***}$ & 0.177$^{***}$ & 0.164$^{***}$ & 0.163$^{***}$ & 0.165$^{***}$ \\ 
  & (0.008) & (0.008) & (0.008) & (0.008) & (0.008) & (0.008) & (0.008) \\ 
 Single family HHs with a lone parent (\%) & 0.094$^{***}$ & 0.095$^{***}$ & 0.091$^{***}$  & 0.093$^{***}$ & 0.044$^{***}$ & 0.039$^{***}$ & 0.080$^{***}$ \\ 
  & (0.011) & (0.011) & (0.011) & (0.011) & (0.011) & (0.011) & (0.011) \\ 
 Socially rented HHs (\%) & 0.138$^{***}$ & 0.137$^{***}$ & 0.137$^{***}$ & 0.140$^{***}$ & 0.157$^{***}$ & 0.165$^{***}$ & 0.145$^{***}$ \\ 
  & (0.010) & (0.011) & (0.010) & (0.010) & (0.010) & (0.010) & (0.010) \\ 
 HHs with 1+ person with a disability (\%) & 0.163$^{***}$ & 0.138$^{***}$ & 0.145$^{***}$ & 0.164$^{***}$ & 0.138$^{***}$ & 0.132$^{***}$ & 0.155$^{***}$\\ 
  & (0.012) & (0.012) & (0.012) & (0.012) & (0.012) & (0.012) & (0.012) \\ 
 Economically inactive (\% of people) & $-$0.052$^{***}$ & $-$0.030$^{***}$ & $-$0.047$^{***}$ & $-$0.052$^{***}$ & $-$0.044$^{***}$ & $-$0.040$^{***}$ & $-$0.054$^{***}$ \\ 
  & (0.010) & (0.010) & (0.010) & (0.010) & (0.010) & (0.010) & (0.010) \\ 
 Part time workers (\% of people) & 0.063$^{***}$ & 0.014 & 0.065$^{***}$ & 0.063$^{***}$ & 0.098$^{***}$ & 0.088$^{***}$ & 0.078$^{***}$ \\ 
  & (0.009) & (0.010) & (0.009) & (0.009) & (0.009) & (0.009) & (0.010) \\ 
 Level 1 qualifications (\% of people) & 0.113$^{***}$ & 0.110$^{***}$ & 0.112$^{***}$ & 0.112$^{***}$ & 0.137$^{***}$ & 0.147$^{***}$ & 0.127$^{***}$  \\
  & (0.011) & (0.011) & (0.011) & (0.011) & (0.010) & (0.011) & (0.011) \\ 
 Level 2 qualifications (\% of people) & $-$0.078$^{***}$ & $-$0.075$^{***}$ & $-$0.078$^{***}$ & $-$0.079$^{***}$ & $-$0.060$^{***}$ & $-$0.057$^{***}$ & $-$0.079$^{***}$ \\ 
  & (0.010) & (0.010) & (0.010) & (0.010) & (0.010) & (0.010) & (0.010) \\ 
 Apprenticeships (\% of people) & $-$0.067$^{***}$ & $-$0.013 & $-$0.061$^{***}$ & $-$0.066$^{***}$ & $-$0.034$^{***}$ & $-$0.027$^{***}$ & $-$0.067$^{***}$ \\ 
  & (0.010) & (0.010) & (0.010) & (0.010) & (0.009) & (0.010) & (0.010) \\ 
 Level 3 qualifications (\% of people) & $-$0.006 & $-$0.010 & $-$0.010 & $-$0.006 & $-$0.019$^{***}$ & $-$0.018$^{***}$ & $-$0.012$^{*}$ \\ 
  & (0.006) & (0.006) & (0.006) & (0.006) & (0.006) & (0.006) & (0.007)  \\ 
 20+ hours of unpaid care (\% of people)  & $-$0.081$^{***}$ & $-$0.066$^{***}$ & $-$0.071$^{***}$ & $-$0.082$^{***}$ & $-$0.098$^{***}$ & $-$0.105$^{***}$ & $-$0.075$^{***}$ \\ 
  & (0.011) & (0.011) & (0.011) & (0.011) & (0.011) & (0.011) & (0.011) \\ 
Larger HHs (\% of people)  & $-$0.030$^{***}$ & 0.023$^{**}$ & $-$0.029$^{***}$ & $-$0.031$^{***}$ & $-$0.048$^{***}$ & $-$0.043$^{***}$ & $-$0.038$^{***}$ \\ 
  & (0.009) & (0.009) & (0.009) & (0.009) & (0.009) & (0.009) & (0.010) \\ 
 Density of the LSOA & $-$0.062$^{***}$ & $-$0.005 & $-$0.055$^{***}$ & $-$0.062$^{***}$ & $-$0.081$^{***}$ & $-$0.094$^{***}$ & $-$0.060$^{***}$ \\ 
  & (0.008) & (0.008) & (0.007) & (0.008) & (0.007) & (0.007) & (0.008) \\ 
 Presence of Trussell centre &  &  & 0.151$^{***}$ &  & 0.123$^{***}$ & 0.116$^{***}$ & 0.147$^{***}$ \\ 
  &  &  & (0.006) &  & (0.006) & (0.006)  & (0.006) \\ 
  Presence of IFAN centre &  &  &  & $-$0.036$^{***}$ & $-$0.027$^{***}$ & $-$0.025$^{***}$ & $-$0.036$^{***}$ \\ 
  &  &  &  & (0.006) & (0.006) & (0.006) & (0.006) \\ 
 Minimum distance  &  &  &  &  & $-$0.220$^{***}$ &  \\ 
  &  &  &  &  & (0.006) &  \\ 
 Minimum public transport travel time &  & &  &  &  & $-$0.235$^{***}$ \\ 
  &  &  &  &  &  & (0.006) \\ 
 Weekly opening hours &  &  &  &  &  &  & 0.037$^{***}$ \\ 
 &  &  &  &  &  &  & (0.007) \\
 Constant & $-$0.000 & 0.416$^{***}$ & 0.000 & $-$0.000 & 0.000 & 0.016$^{***}$ & 0.011$^{*}$\\  
  & (0.006) & (0.126) & (0.006) & (0.006) & (0.005) & (0.006) & (0.006) \\ 

\hline \\[-1.8ex] 
R$^{2}$ & 0.179 & 0.341 & 0.202 & 0.180 & 0.244 & 0.250 & 0.207 \\
Adjusted R$^{2}$ & 0.179 & 0.335 & 0.201 & 0.180 & 0.244 & 0.250 & 0.207 \\
Residual Std. Error & 0.906 & 0.815 & 0.894 & 0.904 & 0.869 & 0.877 & 0.903 \\ 
DF & 25,064 & 25,064 & 25,064 & 25,049 & 25,047 & 24,174 & 23,594\\
\hline \\[-1.8ex] 
\textit{Note:}  & \multicolumn{2}{r}{$^{*}$p$<$0.1; $^{**}$p$<$0.05; $^{***}$p$<$0.01} \\ 
\end{tabular} 
\end{table}

\begin{table}[!t] \centering 
  \caption{Multiple regression models for rural LSOAs, with dependent variable  parcels per 1000 residents, with standardised explanatory variables and accessibility measures.} 
  \label{regression_rural_standardised} 
    \tiny
\begin{tabular}{lccccccc} 
 & \multicolumn{7}{c}{\textit{Dependent variable: Parcels per 1000 residents}} \\ 
\cline{2-8} 
\\[-1.8ex] & R1 & R2 & R3 & R4 & R5 & R6 & R7 \\ 
\hline \\[-1.8ex] 
Universal Credit claimants (\% of people) & 0.165$^{***}$ & 0.205$^{***}$ & 0.150$^{***}$ & 0.170$^{***}$ & 0.156$^{***}$ & 0.157$^{***}$ & 0.161$^{***}$ \\ 
  & (0.011) & (0.011) & (0.011) & (0.012) & (0.011) & (0.012) & (0.013)\\ 
  
Single family HHs with a lone parent (\%) & 0.063$^{***}$ & 0.077$^{***}$ & 0.061$^{***}$ & 0.062$^{***}$ & 0.022 & 0.034$^{*}$ & 0.084$^{***}$ \\ 
  & (0.017) & (0.017) & (0.017) & (0.017) & (0.016) & (0.019) & (0.020) \\ 
  
Socially rented HHs & 0.149$^{***}$ & 0.128$^{***}$ & 0.150$^{***}$ & 0.149$^{***}$ & 0.155$^{***}$ & 0.162$^{***}$ & 0.145$^{***}$ \\  
  & (0.015) & (0.015) & (0.014) & (0.015) & (0.014) & (0.016) & (0.017) \\ 
  
HHs with 1+ person with a disability (\%) & 0.225$^{***}$ & 0.168$^{***}$ & 0.196$^{***}$ & 0.226$^{***}$ & 0.208$^{***}$ & 0.255$^{***}$ & 0.280$^{***}$ \\ 
  & (0.020) & (0.019) & (0.020)  & (0.020) & (0.019) & (0.022) & (0.023) \\ 
  
Economically inactive (\% of people)  & $-$0.095$^{***}$ & $-$0.044$^{***}$ & $-$0.092$^{***}$ & $-$0.095$^{***}$ & $-$0.124$^{***}$ & $-$0.148$^{***}$ & $-$0.128$^{***}$ \\ 
  & (0.018) & (0.017) & (0.018) & (0.018) & (0.017) & (0.019) & (0.021) \\ 
  
Part time workers (\% of people) & 0.066$^{***}$ & $-$0.020 & 0.072$^{***}$ & 0.066$^{***}$ & 0.138$^{***}$ & 0.161$^{***}$  & 0.121$^{***}$  \\ 
  & (0.012) & (0.013) & (0.012) & (0.012) & (0.011) & (0.013) & (0.014) \\ 
  
Level 1 qualifications (\% of people) & 0.014 & 0.046$^{***}$ & 0.017 & 0.013 & 0.023$^{*}$ & 0.047$^{***}$ & 0.035$^{**}$ \\ 
  & (0.013) & (0.014) & (0.013) & (0.013) & (0.012) & (0.015) & (0.016)\\ 
  
Level 2 qualifications (\% of people) & 0.026$^{**}$ & $-$0.002 & 0.026$^{**}$ & 0.025$^{**}$ & 0.039$^{***}$ & 0.043$^{***}$ & 0.029$^{**}$ \\ 
  & (0.011) & (0.010) & (0.011) & (0.011) & (0.010) & (0.012) & (0.013)  \\ 
  
Apprenticeships (\% of people) & $-$0.025$^{**}$ & $-$0.007 & $-$0.014 & $-$0.025$^{**}$ & 0.0004 & $-$0.032$^{***}$ & $-$0.015 \\ 
  & (0.011) & (0.011) & (0.011) & (0.011) & (0.010) & (0.012) & (0.013) \\ 
  
Level 3 qualifications (\% of people) & $-$0.016$^{*}$ & $-$0.028$^{***}$ & $-$0.018$^{*}$ & $-$0.017$^{*}$ & $-$0.038$^{***}$ & $-$0.056$^{***}$ & $-$0.045$^{***}$  \\ 
  & (0.010) & (0.009) & (0.009) & (0.010) & (0.009) & (0.010) & (0.011) \\ 
  
20+ hours of unpaid care (\% of people) & $-$0.120$^{***}$ & $-$0.064$^{***}$ & $-$0.104$^{***}$ & $-$0.120$^{***}$ & $-$0.114$^{***}$ & $-$0.164$^{***}$ & $-$0.162$^{***}$ \\ 
  & (0.016) & (0.015) & (0.016) & (0.016) & (0.015) & (0.017) & (0.019) \\ 
  
Larger HHs (\% of people) & $-$0.004 & 0.036$^{***}$ & 0.003 & $-$0.005 & $-$0.004 & $-$0.001 & $-$0.0004 \\ 
  & (0.012) & (0.011) & (0.011) & (0.012) & (0.011) & (0.012) & (0.013) \\ 
  
Density of the LSOA & 0.017 & 0.044$^{***}$ & 0.017 & 0.017 & $-$0.021$^{**}$ & $-$0.047$^{***}$ & 0.009  \\ 
  & (0.011) & (0.010) & (0.011) & (0.011) & (0.010) & (0.012) & (0.012) \\ 
  
 Presence of Trussell centre &  &  & 0.179$^{***}$ &  & 0.135$^{***}$ & 0.106$^{***}$ & 0.161$^{***}$ \\  
  &  &  & (0.009) &  & (0.008) & (0.009) & (0.010) \\
  
  Presence of IFAN centre &  &  &  & $-$0.030$^{***}$ & $-$0.012 & $-$0.004 & $-$0.022$^{**}$ \\ 
  &  &  &  & (0.009) & (0.008) & (0.011) & (0.011) \\ 
  
 Minimum distance  &  &  &  &  & $-$0.296$^{***}$ &  \\ 
  &  &  &  &  & (0.009) &  \\ 
 Minimum public transport travel time &  & &  &  &  & $-$0.300$^{***}$ \\   
  &  &  &  &  &  & (0.010) \\ 
 Weekly opening hours &  &  &  &  &  &  & $-$0.012 \\ 
 &  &  &  &  &  &  & (0.010) \\
 
 Constant & $-$0.000 & 0.304$^{***}$ & $-$0.000 & $-$0.000 & $-$0.000 & 0.099$^{***}$ & 0.091$^{***}$ \\ 
 & (0.009) & (0.104) & (0.009) & (0.009) & (0.008) & (0.010) & (0.010) \\ 
\hline \\[-1.8ex] 
R$^{2}$& 0.170 & 0.380 & 0.201 & 0.171 & 0.281 & 0.305 & 0.236 \\ 
Adjusted R$^{2}$ & 0.169 & 0.372 & 0.200 & 0.170 & 0.279 & 0.303 & 0.234 \\ 
Residual Std. Error & 0.912 & 0.793 & 0.894 & 0.911 & 0.849 & 0.879 & 0.924 \\ 
DF & 10,594 & 10,474 & 10,593 & 10,593 & 10,591 & 8,271 & 8,122 \\
\hline \\[-1.8ex] 
\textit{Note:}  & \multicolumn{2}{r}{$^{*}$p$<$0.1; $^{**}$p$<$0.05; $^{***}$p$<$0.01} \\ 
\end{tabular} 
\end{table} 

Our findings are consistent with Trussell's Hunger in the UK report \citep{Hunger-in-the-UK} which suggests that those with disabilities are at greater risk of hunger and hardship. This could be as a result of high costs of care and running equipment, which pushes them into hardship. Our findings on universal credit claimants are consistent with research from \citet{Sosenko-2022} who find that increasing universal credit payments by £1 a week would decrease food bank use, on average, by around 120 Trussell food parcels per local authority. 

Turning to negative coefficients, food parcel uptake is lower in areas with a higher share of people who provide over twenty hours of unpaid care a week, are economically inactive, or have apprenticeships or level 2 or 3 qualifications. The largest negative coefficient, a one sd increase in the \% of people providing over twenty hours of unpaid care is associated with a 0.096 sd (or 5.2 parcel) decrease in food parcels.
This is an unexpected result as 2022 research from the Carers Trust suggests that one in seven unpaid carers have used a food bank and one in four say they have to cut back on food as a result of the cost of living crisis \citep{Carers-Trust-2022}.  


To account for unobserved differences between local authorities, we added spatial fixed effects in Model 2. In effect, this looks at whether the relationships seen in Model 1 hold across LSOAs \textit{within} local authorities. 
Two variables - part time workers and population density - become insignificant when the local authority fixed effects are added (population density remains significant, however, in rural areas).  
Some of the fixed effects coefficients (not shown) are significantly larger than the other predictor variables. For example, if the LSOA is in the local authority of Newcastle upon Tyne, this is associated with an increase of 2.7 sd (or 159 parcels) in food parcels per 1000 residents. This corresponds to local hotspots of high food bank use seen in Figure \ref{fig:Fig2}A. The total variance (r2) explained by model 2 climbs to 0.344 (0.335 and 0.372 for urban and rural LSOAs respectively)\footnote{While we drop the fixed effects for the remaining set of models here, we show these models with fixed effects in the SI. There are no significant deviations from the results discussed below.}.


As the main focus of this study, we now investigate a range of accessibility metrics. 
The simplest metric for food bank accessibility is the presence of a Trussell food bank in the LSOA itself. In Model 3 of Table \ref{standardised_regression_results}, unsurprisingly, we show that this is significant with a large coefficient: the presence of a Trussell food bank distribution centre is associated with an increase of 0.16 sd (9.2 parcels) food parcels. This coefficient is similarly high for urban (0.151) and rural (0.179) areas. 

In Model 4, we add a dummy variable for the presence of an IFAN food bank. Since we are investigating Trussell food parcel use, we would expect this to be negative as there should be lower Trussell food parcel uptake in areas in which another centre is located. We do indeed find a negative coefficient that is similar across all three tables: there are on average 2 fewer food parcels per 1000 residents (a decrease of 0.034 sd) distributed compared to if there is was not an IFAN centre in the immediate area. 


Moving to broader accessibility, in Models 5 and 6 we investigate our main explanatory variables of interest: the minimum distance and minimum public transport travel time to a Trussell food bank centre\footnote{The minimum distance and minimum public transport travel time variables are highly correlated and hence we do not include them simultaneously in the same model.}. We control for the presence of a Trussell or IFAN centre in the LSOA, and hence we can interpret the coefficients as capturing the effect of distance/travel time accessibility given that a food bank distribution centre is present (or not). 
Both of these coefficients are large, negative, and statistically significant, suggesting that longer distances and longer travel times are associated with lower food parcel use.
Specifically, we find that a one sd (6.3 km) increase the minimum distance is associated with a 0.234 sd (13.8 parcels) decrease in food parcels. 
Similarly, we find that a one sd (74.8 minutes) increase the minimum travel time is associated with a 0.250 sd (14.8 parcels) decrease in food parcels. 
 
The standardised coefficients are smaller in the urban areas model and larger in the rural areas, suggesting that rural inhabitants are more sensitive to food bank accessibility relative to urban dwellers.
Specifically, we find that, in rural areas, a one sd (8.7 km) increase in minimum distance is associated with a 0.3 sd (17.7 parcels) decrease in food parcels. 
For travel time, an increase of 84 minutes is associated with a decrease of 18 parcels. 

Both of these accessibility variables induce a significant increase in the overall r2, with increases from 0.179 in Model 1 to 0.250 in Model 5 and 0.264 in Model 6.
This effect is particularly strong in rural areas (r2 of 0.308/0.281 respectively) relative to urban areas (r2 of 0.252/0.245 respectively).


Finally, we examine how weekly opening hours (of the closest Trussell food bank) impact food parcel uptake in model 7.  
The coefficient is positive and statistically significant and a one sd (4.2 hours) increase in weekly hours open is associated with a 0.025 (1.5) increase in food parcels.
Although not as large an increase as we see in Models 5 and 6, the opening hours variable increases the overall r2 from 0.2 in Model 3 to 0.209 in Model 7. 
For the urban and rural models, the coefficient for opening times is larger in urban areas than in rural areas. In urban areas, the coefficient is positive and a one sd increase in weekly hours open (3.7 hours) is associated with a 0.037 increase in food parcels (2.2 parcels). In rural areas, the coefficient is negative and not statistically significant. 
The opening hours variable increases the overall r2 for both urban and rural from 0.179 in model U1 to 0.207 in model U7 and 0.169 in model R1 to 0.234 in model R7, respectively.


Overall, we find that accessibility both in terms of public transport travel time and distance to a food bank centre are strong correlates of food parcel use, both in terms of coefficient size and boost to the model predictive power (r2), relative to existing variables from the literature. The effects are even stronger in rural areas where - as we saw above - centres tend not to necessarily be co-located with deprived populations. This latter finding suggests that people in rural settings tend not to access food parcels if distribution centres are not located in close proximity, and that hence there is likely significant unmet need in these areas. Conversely, with shorter opening hours, food parcel uptake is more sensitive to the opening hours of the nearest centre in urban areas. 

\section{\bf{Discussion and conclusion}}

Spatial accessibility to food bank centres is an important, yet under-appreciated, determinant of food bank use in England and Wales. We show that accessibility to food banks in rural areas is worse than their urban counterparts. More subtly, we find that while food banks are located in the most deprived urban areas - with the density of food banks proportional to deprivation level - this is not true in rural areas. Specifically, while we do find food banks in extremely deprived rural areas, they are not consistently located in moderate to mildly deprived areas in which demand is fastest growing. 

Hence, there is a clear need increase accessibility in rural areas through, for example, better public transport access, optimised food bank locations, expanded opening hours or alternative delivery options such as mobile services. There are known inequities between public transport services in urban and rural settings \citep{Sun-2021}, and so for rural dwellers who do not have access to a car, likely to include people with disabilities, older people, and lower income families \citep{Berg-2019}, having a food bank located close to public transport travel hubs is critical. In addition, either optimising current food bank locations or altering the operating model to incorporate a food delivery option (which is already offered in some locations) could better serve the needs of rural communities. This could, for example, include expanding or modifying opening times to include evenings and weekends. 

There are several limitations of this study. Firstly, the main limitation is that we only used Trussell food parcel data and our models do not include parcel data from IFAN centres or from other independent food banks. An interesting avenue for future research is to investigate whether the patterns we find here hold similarly for IFAN (we would expect so). Secondly, we used Euclidean distance between LSOAs and food bank centres rather than measuring distances along the road network, which is likely an underestimation due to, e.g., geographical features such as rivers and lakes. Future work might simulate trips on the road network, or use Google travel times, for slightly finer estimates. Finally, the Trussell food parcel data comes from the tail end of the pandemic (April 2021 to March 2022). While we do not expect this to materially affect our results, the total number of food parcels increased from 2.1 million (21/22) to 2.9 million in 22/23 (and 3.1 million in 23/24). It is conceivable that reluctance to travel due to pandemic effects increased the impact of accessibility on food parcel uptake during this period, which could be investigated in future work. 

\newpage

\bibliographystyle{abbrvnat}
\bibliography{Trussell_references}

\newpage

\section{\bf{Appendix}}

\subsection{Data processing}

The addresses including postcodes, of all the Trussell food bank centres and information about their service model was provided by the Strategic Intelligence Team at Trussell. In the file provided there were 1,642 rows with 10 variables. Once we removed entries that were warehouses and offices and those centres that were temporarily closed, we were left with 1,260 rows. When turning the postcodes provided into point locations, 13 food bank centres were not turned into point locations from using their postcodes. The addresses of all the IFAN food bank centres was downloaded from their website's map function \citep{IFAN_map}. These were both correct as of March 2023. However, we acknowledge that food bank centres will have closed and new ones will have reopened in the time period since. As some food banks in the Trussell Trust Network run a delivery only model, these were removed from the dataset as they do not allow guests to collect food from a fixed location and therefore may distort measures of accessibility. 60 delivery only food bank centres were identified leaving 1,187 centres left as part of our analysis.

The food bank opening times data was matched to the location data of the food bank centres based on their names. However, due to differences in the recorded name for some centres, manual matching was also required.

The addresses and postcodes of all IFAN centres were gathered from their interactive map on their member food bank centres in March 2023 \citep{IFAN_map}. Following the same processing steps for Trussell centres, including removing incompatible postcodes for deriving point locations, a total of 753 IFAN distribution centres remained as part of our analysis.

Data for Universal Credit claimants in England Wales was accessed from the DWP's `Stat-Xplore' site for 2011 LSOAs \citep{DWP-2024}.This required data matching with 2021 LSOAs through spatial lookup files downloaded from the Open Geography Portal \citep{Open-Geog} so that the data was linked to the correct places. Like the parcels per resident variable, a variable called claimants per resident was created and was the number of universal credit claimants as a percentage of the population of the area, using the number of residents in each LSOA from the census.

Spatial lookup files and shape files from the Open Geography Portal \citep{Open-Geog} were used when producing maps of the Local Authority Districts (LADs) and LSOAs.

\subsection{List of variables}

Table \ref{tab:census_variables} and \ref{tab:additional_variables} show the variables used in the econometric models. 

\begin{table}[!ht]
\small
    \centering
    \caption{The variables used from the 2021 Census of England and Wales, all at the scale of 2021 LSOA}
        \begin{tabular}{ll}\toprule
        \textbf{Variable name} & \textbf{Variable Description (\% of households (HHs) or people) } \\ \midrule
        Number of residents &  Number of usual residents living in households in an LSOA \\ 
        Density & Number of usual residents per square kilometre \\
        Household with children & \% of HHs with 1+ child/ren \\ 
        Lone parent families & \% of HHs with 1 parent and 1+ child/ren (of single family HHs) \\
        Socially rented homes & \% of HHs that are socially rented  \\   
        Disabled & \% of HHs with 1+ person who is disabled (the Equality Act) \\
        Bad health & \% of HHs with 1+ person who self-reported “bad” health \\ 
        Very bad health & \% of HHs with 1+ person who self-reported “very bad” health \\ 
        Deprivation & \% of HHs that are deprived in at 2+ domains \\
        Part time workers & \% of workers who work part-time (under 30 hours a week) \\ 
        Economic activity status & \% of people who are economically inactive \\ 
        No qualifications & \% of people with each no qualifications \\
        Level 1 highest qualification & \% of people with level 1 (below a pass at GCSE) \\ 
        Level 2 highest qualification & \% of people with level 2 (pass grade at GCSE) \\
        Level 3 highest qualification & \% of people with level 3 (A level) \\
        Level 4 highest qualification & \% of people with level 4 (university degree) \\
        Household size & \% of people in living a household with 5+ people \\ 
        Unpaid care & \% of people providing 20+ hours of unpaid care a week \\  \bottomrule
        \end{tabular}
    \label{tab:census_variables}
\end{table}

\begin{table}[!ht]
    \small
    \centering
    \caption{The additional sources of data used in this research and modelling}        
    \begin{tabular}{ll}\toprule
        \textbf{Dataset name} & \textbf{Source} \\ \midrule
        Spatial look up files & Open Geography Portal \citep{Open-Geog} \\
        England and Wales 2021 LSOA shape files & Open Geography Portal \citep{Open-Geog} \\ 
        England and Wales 2022 LA shape files & Open Geography Portal \citep{Open-Geog} \\ 
        Addresses of Trussell food bank centres & Trussell Strategic Intelligence Team \\ 
        Addresses of IFAN food bank centres & Independent Food Aid Network \citep{IFAN_map} \\ 
        Universal credit claimants (2011 LSOAs) & \citet{DWP-2024} \\ 
        Urban-ness of areas in England and Wales & OECD Functional Urban Areas \\ 
        LSOA to Functional Urban Area look up file & Open Geography Portal \citep{Open-Geog} \\ \bottomrule
        \end{tabular}
    \label{tab:additional_variables}
\end{table}

\newpage

\subsection{Correlation between variables}

\begin{figure}[!ht]
    \centering
    \includegraphics[width = 13cm]{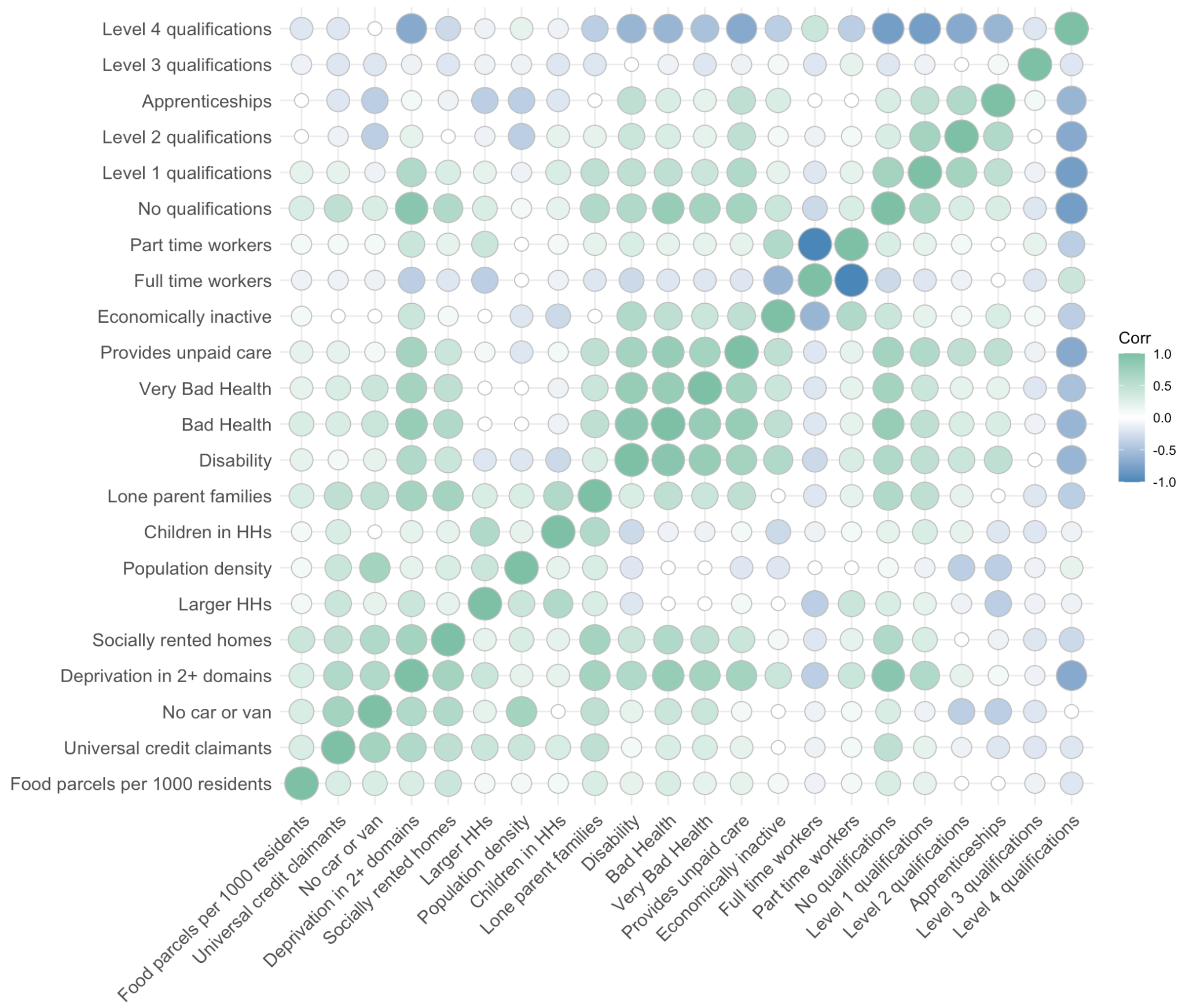}     \caption{Pairwise correlation plot for predictor variables for food bank use. Food parcels per thousand residents is positively correlated with the share of Universal Credit claimants, lone parents families, deprivation and social renting, and having no qualifications. There is high correlation between variables related to health and education, health, disability and education, and Universal Credit claimants, lone parents, no car ownership and social renting.}
    \label{fig:Fig_corr}
\end{figure} 

\newpage

\subsection{Regression Modelling}

The first model used was a step-wise linear regression. Step-wise regression was used as we iteratively added and remove predictors from the regression model in order to find a specific set of predictive variables for the `best performing model'. There are three strategies of step-wise regression:
\begin{enumerate}
    \item \textbf{Forward selection:} to begin there are no predictors in the model and then predictor variables are added until the improvement is no longer statistically significant.
    \item \textbf{Backward selection/elimination:} to begin with all the predictors are in the model and the least `helpful' variables are removed one by one until the model only has variables which are statistically significantly.
    \item \textbf{Step-wise selection:} to begin with there are no predictors in the model and then the most `helpful' predictors are added but when a new variable is added, any that become statistically insignificant then they are removed. 
\end{enumerate} 

Once the `final' variables were defined by the step-wise regression procedure, the four accessibility measures will be added in to the linear regression model. These were replicated for urban areas and non-urban areas.

As well as assessing the statistical significance of each variable added into the regression model, the variance inflation factor (VIF) score was also checked in every iteration of the regression model in order to examine multicollinearity between the variables. When the VIF score for a variable was over 10 (which is a standard threshold value), the variable was removed. 

\begin{table}[!htbp] \centering 
  \caption{Stepwise regression models Table 1} 
  \label{stepwise_regression_1} 
  \tiny
\begin{tabular}{lccccccc} 
 & \multicolumn{7}{c}{\textit{Dependent variable: Parcels per 1000 residents}} \\ 
\cline{2-8} 
\\[-1.8ex] & 1 & 2 & 3 & 4 & 5 & 6 & 7 \\ 
\hline \\[-1.8ex] 
Universal Credit claimants (\% of people) & 8.816$^{***}$ & 5.845$^{***}$ & 5.644$^{***}$ & 4.813$^{***}$ & 4.409$^{***}$ & 4.283$^{***}$ & 4.509$^{***}$ \\ 
  & (0.132) & (0.155) & (0.155) & (0.169) & (0.202) & (0.202) & (0.201) \\ 
 Single family HHs with a lone parent (\%) &  & 2.284$^{***}$ & 2.967$^{***}$ & 2.343$^{***}$ & 2.251$^{***}$ & 1.882$^{***}$ & 0.972$^{***}$ \\ 
  &  & (0.065) & (0.075) & (0.091) & (0.095) & (0.101) & (0.111) \\ 
 HHs with children (\%) &  &  & $-$0.723$^{***}$ & $-$0.534$^{***}$ & $-$0.464$^{***}$ & $-$0.404$^{***}$ & $-$0.337$^{***}$ \\ 
  &  &  & (0.041) & (0.044) & (0.048) & (0.048) & (0.048) \\ 
 Deprivation rank &  &  &  & $-$0.001$^{***}$ & $-$0.001$^{***}$ & $-$0.0002$^{***}$ & $-$0.0003$^{***}$  \\ 
  &  &  &  & (0.00004) & (0.00005) & (0.0001) & (0.0001) \\
  HHs with no car or van (\%) &  &  &  & & 0.104$^{***}$ & 0.082$^{***}$ & $-$0.100$^{***}$ \\ 
  &  &  &  & & (0.028) & (0.029) & (0.030) \\ 
 Deprivation in 2+ domains (\%) &  &  &  & &  & 0.946$^{***}$ & 0.421$^{***}$ \\ 
  &  &  &  & &  & (0.093) & (0.096)  \\ 
 Socially rented HHs (\%) &  &  &  & &  &  & 0.641$^{***}$ \\ 
  &  &  &  & &  &  & (0.032) \\ 
 Constant & 8.857$^{***}$ & $-$9.123$^{***}$ & 4.570$^{***}$ & 16.640$^{***}$ & 14.070$^{***}$ & $-$0.502 & 8.226$^{***}$ \\ 
  & (0.448) & (0.674) & (1.031) & (1.439) & (1.601) & (2.142) & (2.173) \\ 
\hline \\[-1.8ex] 
Observations & 35,672 & 35,672 & 35,672 & 35,672 & 35,672 & 35,672 & 35,672 \\ 
R$^{2}$ & 0.111 & 0.140 & 0.148 & 0.151 & 0.152 & 0.154 & 0.163 \\ 
Adjusted R$^{2}$ & 0.111 & 0.140 & 0.148 & 0.151 & 0.151 & 0.154 & 0.163 \\ 
Residual Std. Error & 55.617 & 54.676 & 54.443 & 54.335 & 54.325 & 54.246 & 53.942  \\ 
\hline \\[-1.8ex] 
\textit{Note:}  & \multicolumn{4}{r}{$^{*}$p$<$0.1; $^{**}$p$<$0.05; $^{***}$p$<$0.01} \\ 
\end{tabular} 
\end{table}

\begin{table}[!htbp] \centering 
  \caption{Stepwise regression models Table 2} 
  \label{stepwise_regression_2} 
  \tiny
\begin{tabular}{lccccccc} 
 & \multicolumn{7}{c}{\textit{Dependent variable: Parcels per 1000 residents}} \\ 
\cline{2-8} 
\\[-1.8ex] & 8 & 9 & 10 & 11 & 12 & 13 & 14 \\ 
\hline \\[-1.8ex] 
 Universal Credit claimants (\% of people) & 4.767$^{***}$ & 4.794$^{***}$ & 4.756$^{***}$ & 4.753$^{***}$ & 4.620$^{***}$ & 4.810$^{***}$ & 4.726$^{***}$\\ 
  & (0.202) & (0.202) & (0.202) & (0.202) & (0.201) & (0.197) & (0.197)  \\ 
 Single family HHs with a lone parent (\%) & 0.730$^{***}$ & 0.709$^{***}$ & 0.710$^{***}$ & 0.650$^{***}$ & 0.551$^{***}$ & 0.664$^{***}$ & 0.576$^{***}$ \\ 
  & (0.112) & (0.112) & (0.112) & (0.114) & (0.114) & (0.094) & (0.095)  \\ 
 HHs with children (\%) & 0.131$^{**}$ & 0.151$^{**}$ & 0.151$^{**}$ & 0.132$^{**}$ & 0.016 \\ 
 & (0.062) & (0.063) & (0.063) & (0.063) & (0.061) \\ 
 Deprivation rank & $-$0.0001$^{**}$  & $-$0.0001$^{***}$ & $-$0.0001$^{***}$ & $-$0.0001$^{**}$ & $-$0.0001 \\ 
 & (0.0001) & (0.0001) & (0.0001) & (0.0001) & (0.0001) \\ 
 HHs with no car or van (\%) & 0.092$^{***}$ & 0.103$^{***}$ & 0.105$^{***}$ & 0.093$^{***}$ & 0.082$^{**}$ & 0.076$^{***}$ & 0.160$^{***}$ \\ 
 & (0.034) & (0.034) & (0.034) & (0.034) & (0.033) & (0.028) & (0.031) \\ 
 Deprivation in 2+ domains (\%) & $-$0.379$^{***}$ & $-$0.253$^{**}$ & $-$0.320$^{***}$ & $-$0.204$^{*}$ &  \\ 
  & (0.117) & (0.122) & (0.118) & (0.124) &  \\ 
 Socially rented HHs (\%) & 0.587$^{***}$ & 0.589$^{***}$ & 0.592$^{***}$ & 0.593$^{***}$ & 0.581$^{***}$ & 0.579$^{***}$ & 0.568$^{***}$ \\ 
  & (0.032) & (0.032) & (0.032) & (0.032) & (0.032) & (0.032) & (0.032) \\ 
 HHs with 1+ person with a disability (\%) & 1.356$^{***}$ & 1.750$^{***}$ & 1.554$^{***}$ & 1.646$^{***}$ & 1.706$^{***}$ & 1.774$^{***}$ & 1.599$^{***}$ \\ 
  & (0.161) & (0.130) & (0.133) & (0.126) & (0.105) & (0.108) & (0.115) \\ 
 Bad health (\% of people) & & $-$1.538$^{***}$ &  &  &  \\ 
  & & (0.437) &  &  &  \\ 
 Very bad health (\% of people) & &  & $-$2.426$^{***}$ & $-$2.292$^{***}$ & $-$1.763$^{**}$ & $-$1.245$^{*}$  \\ 
  & &  & (0.739) & (0.740) & (0.737) & (0.744) \\ 
 Economically inactive (\% of people) & &  &  & $-$0.150$^{***}$ & $-$0.553$^{***}$ & $-$0.537$^{***}$ & $-$0.438$^{***}$ \\ 
  & &  &  & (0.048) & (0.061) & (0.062) & (0.064) \\ 
 Part time workers (\% of people) & &  &  &  & 0.686$^{***}$ & 0.696$^{***}$ & 0.656$^{***}$  \\ 
  & &  &  &  & (0.074) & (0.073) & (0.073) \\ 
No qualifications (\% of people) & &  &  &  & & $-$0.143$^{**}$ & $-$0.440$^{***}$ \\ 
  & &  &  &  & & (0.067) & (0.081) \\ 
 Level 1 qualifications (\% of people) & &  &  &  & &  & 1.101$^{***}$ \\ 
  & &  &  &  & &  & (0.170) \\ 
 Constant & $-$21.398$^{***}$ & $-$24.428$^{***}$ & $-$23.381$^{***}$ & $-$19.816$^{***}$ & $-$25.044$^{***}$ & $-$27.194$^{***}$ & $-$33.458$^{***}$ \\ 
   & (3.320) & (3.430) & (3.374) & (3.561) & (3.584) & (2.013) & (2.233)  \\ 
\hline \\[-1.8ex] 
Observations & 35,672 & 35,672 & 35,672 & 35,672 & 35,672 & 35,672 \\ 
R$^{2}$ & 0.167 & 0.167 & 0.167 & 0.167 & 0.169 & 0.169 & 0.170 \\ 
Adjusted R$^{2}$ & 0.167 & 0.167 & 0.167 & 0.167 & 0.169 & 0.169 & 0.172 \\ 
Residual Std. Error & 53.838 & 53.829 & 53.830 & 53.824 & 53.761 \\ 
\hline \\[-1.8ex] 
\textit{Note:}  & \multicolumn{4}{r}{$^{*}$p$<$0.1; $^{**}$p$<$0.05; $^{***}$p$<$0.01} \\ 
\end{tabular} 
\end{table}

\begin{table}[!htbp] \centering 
  \caption{Stepwise regression models Table 3} 
  \label{stepwise_regression_3} 
  \tiny
\begin{tabular}{lccccccc} 
 & \multicolumn{7}{c}{\textit{Dependent variable: Parcels per 1000 residents}} \\ 
\cline{2-8} 
\\[-1.8ex] & 15 & 16 & 17 & 18 & 19 & 20 & 21 \\ 
\hline \\[-1.8ex] 
 Universal Credit claimants (\% of people) & 4.739$^{***}$ & 4.592$^{***}$ & 4.750$^{***}$ & 4.690$^{***}$ & 4.446$^{***}$ & 4.483$^{***}$ & 4.633$^{***}$ \\ 
  & (0.197) & (0.198) & (0.177) & (0.179) & (0.179) & (0.180) & (0.175) \\ 
 Single family HHs with a lone parent (\%) & 0.658$^{***}$ & 0.591$^{***}$ & 0.609$^{***}$ & 0.614$^{***}$ & 0.920$^{***}$ & 0.932$^{***}$ & 1.012$^{***}$ \\ 
 & (0.096) & (0.097) & (0.097) & (0.097) & (0.101) & (0.101) & (0.101) \\ 
 HHs with no car or van (\%) & 0.083$^{**}$ & 0.061$^{*}$ \\ 
 & (0.035) & (0.035) \\ 
 Socially rented HHs (\%) & 0.564$^{***}$ & 0.525$^{***}$ & 0.541$^{***}$ & 0.547$^{***}$ & 0.508$^{***}$ & 0.502$^{***}$ & 0.512$^{***}$ \\ 
 & (0.032) & (0.032) & (0.031) & (0.031) & (0.031) & (0.031) & (0.031) \\ 
 HHs with 1+ person with a disability (\%) & 1.697$^{***}$ & 1.922$^{***}$ & 1.969$^{***}$ & 2.032$^{***}$ & 2.422$^{***}$ & 2.225$^{***}$ & 2.217$^{***}$ \\ 
  & (0.098) & (0.103) & (0.100) & (0.103) & (0.109) & (0.129) & (0.123) \\ 
 Economically inactive (\% of people) & $-$0.470$^{***}$ & $-$0.456$^{***}$ & $-$0.465$^{***}$ & $-$0.496$^{***}$ & $-$0.378$^{***}$ & $-$0.372$^{***}$ & $-$0.458$^{***}$  \\ 
  & (0.064) & (0.064) & (0.064) & (0.065) & (0.064) & (0.064) & (0.063) \\ 
 Part time workers (\% of people) & 0.666$^{***}$ & 0.606$^{***}$ & 0.617$^{***}$ & 0.620$^{***}$ & 0.562$^{***}$ & 0.641$^{***}$ & 0.654$^{***}$ \\ 
  & (0.073) & (0.074) & (0.076) & (0.076) & (0.076) & (0.080) & (0.080) \\ 
No qualifications (\% of people) & $-$0.478$^{***}$ & $-$0.437$^{***}$ & $-$0.447$^{***}$ & $-$1.574$^{***}$ & $-$0.210$^{**}$ & $-$0.106  \\ 
 & (0.078) & (0.078) & (0.082) & (0.400) & (0.085) & (0.092) \\ 
 Level 1 qualifications (\% of people) & 1.492$^{***}$ & 1.838$^{***}$ & 1.812$^{***}$ & 0.628 & 1.817$^{***}$ & 1.878$^{***}$ & 1.792$^{***}$ \\ 
 & (0.188) & (0.194) & (0.193) & (0.454) & (0.192) & (0.194) & (0.180) \\ 
 Level 2 qualifications (\% of people) & $-$0.751$^{***}$ & $-$0.564$^{***}$ & $-$0.693$^{***}$ & $-$1.701$^{***}$ & $-$0.515$^{***}$ & $-$0.544$^{***}$ & $-$0.872$^{***}$ \\ 
 & (0.167) & (0.168) & (0.150) & (0.381) & (0.151) & (0.151) & (0.157) \\ 
 Apprenticeships (\% of people) &  & $-$1.614$^{***}$ & $-$1.623$^{***}$ & $-$2.716$^{***}$ & $-$1.417$^{***}$ & $-$1.662$^{***}$ & $-$1.776$^{***}$ \\ 
 &  & (0.224) & (0.224) & (0.441) & (0.225) & (0.240) & (0.235) \\ 
 Level 3 qualifications (\% of people) & & & $-$0.065 & $-$1.111$^{***}$ & $-$0.133$^{*}$ & $-$0.111 & $-$0.089 \\ 
  & & & (0.068) & (0.370) & (0.068) & (0.069) & (0.065)\\ 
 Level 4 qualifications (\% of people) & & & & $-$1.076$^{***}$ &  &  \\ 
  & & &  & (0.374) &  &  \\ 
 20+ hours of unpaid care (\% of people) & & &  &  & $-$3.573$^{***}$ & $-$3.387$^{***}$ & $-$3.578$^{***}$ \\ 
  & & &  &  & (0.336) & (0.342) & (0.337) \\ 
 Living in larger HHs (\% of people) &  &  &  & & & $-$0.260$^{***}$ & $-$0.204$^{**}$ \\ 
  &  &  &  & & & (0.090) & (0.084) \\ 
  Density of LSOA & & & & & & & $-$0.001$^{***}$  \\
  & & & & & & & (0.0001) \\
 Constant & $-$26.949$^{***}$ & $-$25.252$^{***}$ & $-$22.268$^{***}$ & 83.048$^{**}$ & $-$24.633$^{***}$ & $-$24.016$^{***}$ & $-$15.675$^{***}$ \\ 
 & (2.608) & (2.616) & (2.238) & (36.657) & (2.245) & (2.255) & (2.458) \\ 
\hline \\[-1.8ex] 
Observations & 35,672 & 35,672 & 35,672 & 35,672 & 35,672 & 35,672 & 35,672 \\ 
R$^{2}$ & 0.171 & 0.172 & 0.172 & 0.172 & 0.174 & 0.175 & 0.176 \\ 
Adjusted R$^{2}$ & 0.170 & 0.172 & 0.172 & 0.172 & 0.174 & 0.174 & 0.176 \\ 
Residual Std. Error & 53.712 & 53.674 & 53.676 & 53.670 & 53.591 & 53.586 & 53.545\\ 
\hline \\[-1.8ex] 
\textit{Note:}  & \multicolumn{4}{r}{$^{*}$p$<$0.1; $^{**}$p$<$0.05; $^{***}$p$<$0.01} \\ 
\textit{Note:}  & \multicolumn{4}{r}{Level 4 qualifications variable was removed due to very high VIF scores} \\ 
\end{tabular} 
\end{table}

\begin{table}[!t] \centering 
  \caption{Multiple regression models for predicting food bank use with standardised predictor variables and accessibility measures added and fixed effects} 
  \label{standardised_regression_results_fe}
  \tiny
\begin{tabular}{lccccc} 
 & \multicolumn{5}{c}{\textit{Dependent variable: Parcels per 1000 residents}} \\ 
\cline{2-6} 
\\[-1.8ex] & FE3 & FE4 & FE5 & FE6 & FE7 \\ 
\hline \\[-1.8ex] 
 Universal Credit claimants (\% of people) & 0.220$^{***}$ & 0.235$^{***}$ & 0.216$^{***}$ & 0.212$^{***}$ & 0.221$^{***}$ \\ 
  & (0.006) & (0.006) & (0.006) & (0.007) & (0.007) \\ 
  
 Single family HHs with a lone parent (\%) & 0.091$^{***}$ & 0.094$^{***}$ & 0.078$^{***}$ & 0.081$^{***}$ & 0.092$^{***}$  \\ 
  & (0.009) & (0.009) & (0.009) & (0.010) & (0.010)  \\ 
  
 Socially rented HHs (\%) & 0.138$^{***}$ & 0.136$^{***}$ & 0.144$^{***}$ & 0.153$^{***}$ & 0.146$^{***}$ \\ 
    & (0.009) & (0.009) & (0.009) & (0.009) & (0.009) \\ 
  
 HHs with 1+ person with a disability (\%) & 0.133$^{***}$ & 0.150$^{***}$ & 0.129$^{***}$ & 0.135$^{***}$ & 0.150$^{***}$ \\ 
  & (0.010) & (0.010) & (0.010) & (0.011) & (0.011) \\ 
  
 Economically inactivity (\% of people) & $-$0.031$^{***}$ & $-$0.033$^{***}$ & $-$0.036$^{***}$ & $-$0.036$^{***}$ & $-$0.038$^{***}$ \\ 
  & (0.008) & (0.009) & (0.008) & (0.009) & (0.009) \\ 
  
 Part time workers (\% of people) & 0.005 & 0.001 & 0.025$^{***}$ & 0.019$^{**}$ & 0.003 \\ 
 & (0.008) & (0.008) & (0.008) & (0.008) & (0.009) \\ 
  
 Level 1 qualification (\% of people) & 0.093$^{***}$ & 0.094$^{***}$ & 0.094$^{***}$ & 0.106$^{***}$ & 0.115$^{***}$ \\ 
 & (0.009) & (0.009) & (0.009) & (0.009) & (0.010) \\ 
  
 Level 2 qualification (\% of people) & $-$0.050$^{***}$ & $-$0.051$^{***}$ & $-$0.046$^{***}$ & $-$0.048$^{***}$ & $-$0.056$^{***}$\\ 
  & (0.008) & (0.008) & (0.008) & (0.008) & (0.009)\\ 
  
 Apprenticeships (\% of people) & $-$0.010 & $-$0.013$^{*}$ & $-$0.012 & $-$0.024$^{***}$ & $-$0.024$^{***}$ \\ 
  & (0.008) & (0.008) & (0.008) & (0.008) & (0.008) \\ 
  
 Level 3 qualification (\% of people) & $-$0.013$^{**}$ & $-$0.011$^{**}$ & $-$0.020$^{***}$ & $-$0.020$^{***}$ & $-$0.011$^{**}$ \\ 
  & (0.005) & (0.005) & (0.005) & (0.005) & (0.005) \\ 
  
 20 + hours of unpaid care (\% of people) & $-$0.060$^{***}$ & $-$0.070$^{***}$ & $-$0.062$^{***}$ & $-$0.072$^{***}$ & $-$0.071$^{***}$ \\ 
  & (0.009) & (0.009) & (0.009) & (0.009) & (0.010) \\ 
  
 Larger HHs (\% of people) & 0.030$^{***}$ & 0.030$^{***}$ & 0.019$^{***}$ & 0.020$^{***}$ & 0.028$^{***}$  \\ 
  & (0.007) & (0.007) & (0.007) & (0.008) & (0.008) \\ 
  
 Density of the LSOA & 0.007 & 0.005 & $-$0.005 & $-$0.014$^{**}$ & 0.008 \\ 
  & (0.007) & (0.007) & (0.007) & (0.007) & (0.007) \\ 
  
 Presence of Trussell centre & 0.119$^{***}$ &  & 0.103$^{***}$ & 0.094$^{***}$ & 0.114$^{***}$ \\ 
  & (0.004) &  & (0.004) & (0.004) & (0.005) \\ 
  
 Presence of IFAN centre &  & $-$0.017$^{***}$ & $-$0.015$^{***}$ & $-$0.011$^{**}$ & $-$0.016$^{***}$ \\ 
 &  & (0.004) & (0.004) & (0.005) & (0.005) \\ 
 Minimum distance &  &  & $-$0.219$^{***}$\\ 
 &  &  & (0.007) \\ 
 Minimum public transport travel time & & & & $-$0.216$^{***}$ & \\ 
 & & & & (0.007) & \\
 Weekly opening hours &  &  &  &  & 0.021$^{***}$ \\ 
 &  &  &  &  & (0.007) \\
 Constant & 0.428$^{***}$ & 0.421$^{***}$ & 0.300$^{**}$ & 0.313$^{**}$ & 0.427$^{***}$ \\ 
 & (0.124) & (0.125) & (0.123) & (0.125) & (0.127) \\ 
\hline \\[-1.8ex] 
Fixed effects & Yes & Yes & Yes & Yes & Yes \\
R$^{2}$ & 0.364 & 0.351 & 0.380 & 0.384 & 0.370  \\ 
Adjusted R$^{2}$ & 0.358 & 0.344 & 0.374 & 0.378 & 0.364 \\ 
Residual Std. Error & 0.801 & 0.810 & 0.792 & 0.809 & 0.819 \\ 
DF & 35,672 & 35,672 & 35,672 & 32,479 & 31,716 \\ 
\hline \\[-1.8ex] 
\textit{Note:}  & \multicolumn{2}{r}{$^{*}$p$<$0.1; $^{**}$p$<$0.05; $^{***}$p$<$0.01} \\ 
\end{tabular} 
\end{table}


We also experimented with more complex accessibility models, such as two step floating catchment models \citep{Chen-2019}, but they were significantly less predictive than simple time and distance based metrics. 

\begin{figure}[!t]
    \centering
    \includegraphics[width = 14cm]{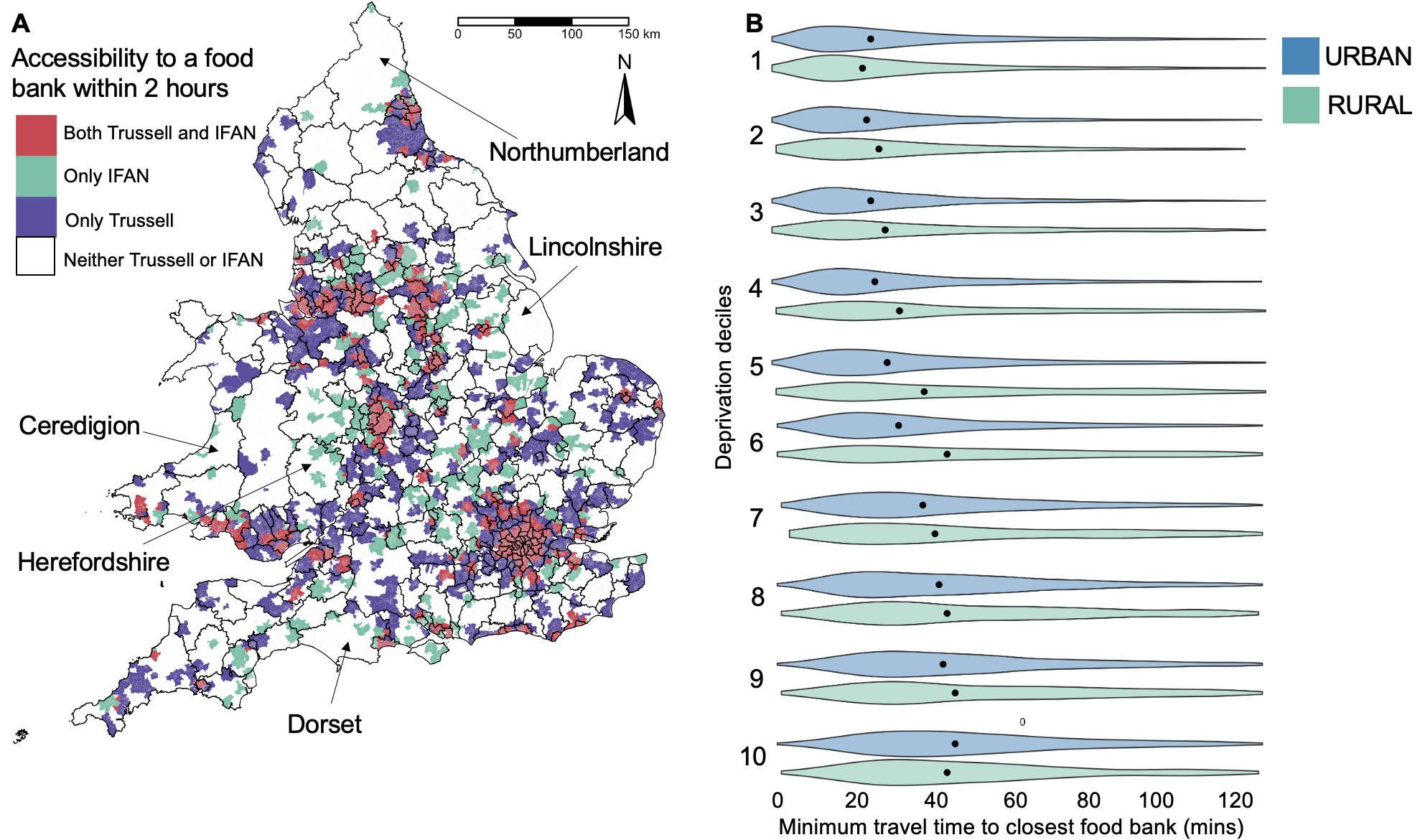}
    \caption{}
    \label{fig:Fig3_appendix}
\end{figure}

The coefficient for minimum distance is an order of magnitude larger than travel time: a 1 km increase in a distance is associated with a decrease of 2.2 food parcels per 1000 residents, while a 1 minute increase in the travel time is associated with a decrease of 0.2 food parcels per 1000 residents. For a 10 minute increase in minimum travel time, there is a decrease of 2.0 food parcels per 1000 residents, which is similar to the decrease when we have a 1km increase in distance. 

In urban areas, a 1 km increase in distance is associated with a decrease of 2.9 food parcels per 1000 residents, while a 1 minute increase in the travel time is associated with a decrease of 0.21 food parcels per 1000 residents. And for rural areas, a 1 km increase in distance is associated with a decrease of 1.8 food parcels per 1000 residents, while a 1 minute increase in the travel time is associated with a decrease of 0.19 food parcels per 1000 residents. This suggests that food bank users in rural areas travel longer distances to access food parcels compare to residents of urban areas.  
\begin{table}[!t] \centering 
  \caption{Multiple regression models for predicting food bank use with predictor variables and accessibility measures added} 
  \label{regression_rural} 
    \tiny
\begin{tabular}{@{\extracolsep{5pt}}lccccccc} 
 & \multicolumn{7}{c}{\textit{Dependent variable: Parcels per 1000 residents}} \\ 
\cline{2-8} 
\\[-1.8ex] & (1) & (2) & (3) & (4) & (5) & (6) & (7) \\ 
\hline \\[-1.8ex] 
 \% of people who claim Universal Credit & 4.633$^{***}$ & 6.155$^{***}$ & 4.307$^{***}$ & 4.751$^{***}$ & 4.336$^{***}$ & 4.262$^{***}$ & 4.358$^{***}$ \\ 
  & (0.175) & (0.170) & (0.173) & (0.176) & (0.169) & (0.175) & (0.182)  \\ 
 \% of single family HHs with a lone parent & 1.012$^{***}$ & 1.039$^{***}$ & 0.986$^{***}$ & 1.000$^{***}$ & 0.398$^{***}$ & 0.318$^{***}$ & 0.836$^{***}$ \\ 
  & (0.101) & (0.102) & (0.099) & (0.101) & (0.097) & (0.102) & (0.106) \\ 
 \% of HHs that are socially rented & 0.512$^{***}$ & 0.500$^{***}$ & 0.514$^{***}$ & 0.517$^{***}$ & 0.581$^{***}$ & 0.619$^{***}$ & 0.535$^{***}$ \\ 
  & (0.031) & (0.032) & (0.030) & (0.031) & (0.029) & (0.031) & (0.033) \\ 
 \% of HHs with 1+ person with a disability & 2.217$^{***}$ & 1.793$^{***}$ & 1.946$^{***}$ & 2.224$^{***}$ & 1.959$^{***}$ & 1.980$^{***}$ & 2.289$^{***}$ \\ 
  & (0.123) & (0.123) & (0.122) & (0.123) & (0.118) & (0.126) & (0.132) \\ 
 \% people who are economically inactive & $-$0.458$^{***}$ & $-$0.240$^{***}$ & $-$0.429$^{***}$ & $-$0.460$^{***}$ & $-$0.502$^{***}$ & $-$0.481$^{***}$ & $-$0.527$^{***}$ \\ 
  & (0.063) & (0.062) & (0.062) & (0.063) & (0.061) & (0.064) & (0.068) \\ 
  \% people who work part time & 0.654$^{***}$ & 0.014 & 0.688$^{***}$ & 0.655$^{***}$ & 1.203$^{***}$ & 1.148$^{***}$ & 0.931$^{***}$\\ 
  & (0.080) & (0.087) & (0.079) & (0.080) & (0.077) & (0.083) & (0.087) \\ 
 \% people with level 1 qualifications & 1.792$^{***}$ & 1.993$^{***}$ & 1.789$^{***}$ & 1.789$^{***}$ & 2.154$^{***}$ & 2.686$^{***}$ & 2.370$^{***}$ \\ 
  & (0.180) & (0.191) & (0.177) & (0.180) & (0.172) & (0.184) & (0.193) \\ 
 \% people with level 2 qualifications & $-$0.872$^{***}$ & $-$1.017$^{***}$ & $-$0.888$^{***}$ & $-$0.880$^{***}$ & $-$0.454$^{***}$ & $-$0.403$^{**}$ & $-$0.919$^{***}$ \\ 
  & (0.157) & (0.161) & (0.155) & (0.157) & (0.150) & (0.161) & (0.168) \\ 
 \% people with apprenticeships & $-$1.776$^{***}$ & $-$0.405$^{*}$ & $-$1.527$^{***}$ & $-$1.775$^{***}$ & $-$0.954$^{***}$ & $-$1.114$^{***}$ & $-$1.857$^{***}$ \\ 
  & (0.235) & (0.241) & (0.232) & (0.235) & (0.225) & (0.242) & (0.252)  \\ 
 \% people with level 3 qualifications & $-$0.089 & $-$0.134$^{**}$ & $-$0.128$^{**}$ & $-$0.085 & $-$0.264$^{***}$ & $-$0.281$^{***}$ & $-$0.213$^{***}$ \\ 
  & (0.065) & (0.065) & (0.064) & (0.065) & (0.062) & (0.065) & (0.068) \\ 
  \% people provide 20+ hours unpaid care & $-$3.578$^{***}$ & $-$2.584$^{***}$ & $-$3.120$^{***}$ & $-$3.611$^{***}$ & $-$3.722$^{***}$ & $-$4.463$^{***}$ & $-$3.716$^{***}$ \\ 
  & (0.337) & (0.333) & (0.333) & (0.337) & (0.323) & (0.346) & (0.362) \\ 
 \% people in living in larger HHs & $-$0.204$^{**}$ & 0.341$^{***}$ & $-$0.181$^{**}$ & $-$0.213$^{**}$ & $-$0.417$^{***}$ & $-$0.408$^{***}$ & $-$0.302$^{***}$ \\ 
  & (0.084) & (0.083) & (0.082) & (0.084) & (0.080) & (0.084) & (0.087) \\ 
 Density of the LSOA & $-$0.001$^{***}$ & 0.0001 & $-$0.001$^{***}$ & $-$0.001$^{***}$ & $-$0.001$^{***}$ & $-$0.001$^{***}$ & $-$0.001$^{***}$ \\ 
   & (0.0001) & (0.0001) & (0.0001) & (0.0001) & (0.0001) & (0.0001) & (0.0001) \\ 
 Presence of TT food bank & & & 50.015$^{***}$ & & 40.944$^{***}$ & 36.284$^{***}$ & 47.937$^{***}$ \\ 
  & & & (1.508) &  & (1.475) & (1.528) & (1.582) \\ 
 Presence of IFAN food bank & & &  & $-$15.114$^{***}$ & $-$10.299$^{***}$ & $-$9.127$^{***}$ & $-$14.616$^{***}$ \\ 
  & & &  & (2.184) & (2.090) & (2.280) & (2.381) \\ 
Minimum distance to TT food bank & & & & & $-$2.204$^{***}$ &  &  \\ 
  & & & & & (0.047) &  &  \\ 
 Minimum public transport travel time & & & & &  & $-$0.197$^{***}$ &  \\ 
  & & & & &  & (0.004) &  \\ 
 Total weekly opening hours & & & & &  &  & 0.338$^{***}$ \\ 
  & & & & &  &  & (0.072) \\ 
 Constant & $-$15.675$^{***}$ & 5.531 & $-$16.795$^{***}$ & $-$15.411$^{***}$ & $-$16.395$^{***}$ & $-$8.451$^{***}$ & $-$22.519$^{***}$ \\ 
  & (2.458) & (8.090) & (2.422) & (2.457) & (2.349) & (2.481) & (2.606) \\ 
\hline \\[-1.8ex] 
Fixed effects & No & Yes & No & No & No & No & No \\
R$^{2}$ & 0.176 & 0.350 & 0.201 & 0.177 & 0.248 & 0.259 & 0.210 \\ 
Adjusted R$^{2}$ & 0.176 & 0.344 & 0.200 & 0.177 & 0.248 & 0.258 & 0.209 \\ 
Residual Std. Error & 53.545 & 47.758 & 52.739 & 53.510 & 51.144 & 52.080 & 53.851 \\ 
DF & 35658 & 35328 & 35657 & 35657 & 35655 & 32462 & 31699 \\ 
\hline \\[-1.8ex] 
\textit{Note:}  & \multicolumn{2}{r}{$^{*}$p$<$0.1; $^{**}$p$<$0.05; $^{***}$p$<$0.01} \\ 
\end{tabular} 
\end{table} 

\begin{table}[!htbp] \centering 
  \caption{Multiple regression model for predicting food bank use with predictor variables and the two-step floating area catchment model} 
  \label{regression_fca} 

\begin{tabular}{@{\extracolsep{5pt}}lc} 
& \multicolumn{1}{c}{\textit{Dependent variable: Parcels per 1000 residents}} \\ 
\cline{2-2} 
\\[-1.8ex] & (8) \\ 
\hline \\[-1.8ex] 
\% of people who claim Universal Credit & 0.005$^{***}$ \\ 
  & (0.0002) \\ 
\% of single family HHs with a lone parent (\%) & 0.002$^{***}$ \\ 
  & (0.0001) \\ 
\% of HHs that are socially rented & 0.003$^{***}$ \\ 
  & (0.0001) \\ 
\% of HHs with 1+ person with a disability & $-$0.0004$^{***}$ \\ 
  & (0.0001) \\ 
\% people who are economically inactive & 0.001$^{***}$  \\ 
  & (0.0001) \\ 
 \% people who work part time & 0.002$^{***}$ \\ 
  & (0.0002) \\ 
\% people with level 1 qualifications & $-$0.001$^{***}$ \\ 
  & (0.0002) \\ 
\% people with level 2 qualifications & $-$0.002$^{***}$ \\ 
  & (0.0002) \\ 
\% people with apprenticeships & $-$0.0002$^{***}$ \\ 
  & (0.0001) \\ 
\% people with level 3 qualifications & $-$0.004$^{***}$ \\ 
  & (0.0003) \\ 
\% people provide 20+ hours unpaid care & $-$0.0003$^{***}$ \\ 
  & (0.0001) \\ 
 \% people in living in larger HHs & $-$0.00000$^{***}$ \\ 
  & (0.00000) \\ 
 Density of the LSOA & 0.004$^{***}$ \\ 
  & (0.0003) \\ 
 Two-step floating catchment area value & $-$0.022$^{***}$  \\ 
  & (0.002) \\
\hline \\[-1.8ex] 
Fixed effects & No \\
R$^{2}$ & 0.174 \\ 
Adjusted R$^{2}$ & 0.174 \\ 
Residual Std. Error & 0.053 \\ 
DF & 35473) \\ 
\hline 
\hline \\[-1.8ex] 
\textit{Note:}  & \multicolumn{1}{r}{$^{*}$p$<$0.1; $^{**}$p$<$0.05; $^{***}$p$<$0.01} \\ 
\end{tabular} 
\end{table}

\begin{table}[!t] \centering 
  \caption{Multiple regression models for predicting food bank use in urban areas with predictor variables and accessibility measures added} 
  \label{regression_urban_unstandardised} 
    \tiny
\begin{tabular}{@{\extracolsep{5pt}}lcccccc} 
 & \multicolumn{6}{c}{\textit{Dependent variable: Parcels per 1000 residents}} \\ 
\cline{2-7} 
\\[-1.8ex] & U1 & U2 & U3 & U4 & U5 & U6 \\ 
\hline \\[-1.8ex] 
 \% of people who claim Universal Credit & 4.325$^{***}$ & 6.012$^{***}$ & 4.056$^{***}$ & 4.437$^{***}$ & 4.016$^{***}$ & 3.991$^{***}$ \\ 
  & (0.206) & (0.203) & (0.203) & (0.207) & (0.198) & (0.201) \\ 
 \% of single family HHs with a lone parent & 0.585$^{***}$ & 0.640$^{***}$ & 0.580$^{***}$ & 0.574$^{***}$ & 0.131 & 0.021 \\ 
  & (0.125) & (0.128) & (0.123) & (0.125) & (0.121) & (0.123) \\ 
 \% of people who are students  & 1.060$^{***}$ & 0.995$^{***}$ & 1.004$^{***}$ & 1.054$^{***}$ & 0.835$^{***}$ & 0.977$^{***}$ \\ 
  & (0.107) & (0.108) & (0.106) & (0.107) & (0.103) & (0.106) \\ 
 \% of HHs that are socially rented & 0.390$^{***}$ & 0.374$^{***}$ & 0.393$^{***}$ & 0.398$^{***}$ & 0.473$^{***}$ & 0.493$^{***}$ \\ 
  & (0.038) & (0.040) & (0.037) & (0.038) & (0.036) & (0.037) \\ 
 \% of HHs with 1+ person with a disability & 2.489$^{***}$ & 2.283$^{***}$ & 2.237$^{***}$ & 2.489$^{***}$ & 2.093$^{***}$ & 2.067$^{***}$ \\ 
  & (0.153) & (0.158) & (0.151) & (0.153) & (0.147) & (0.152) \\ 
 \% people who are economically inactive & $-$0.607$^{***}$ & $-$0.447$^{***}$ & $-$0.560$^{***}$ & $-$0.607$^{***}$ & $-$0.507$^{***}$ & $-$0.505$^{***}$ \\ 
  & (0.079) & (0.078) & (0.078) & (0.079) & (0.076) & (0.078) \\ 
 \% people who work part time & 0.643$^{***}$ & 0.065 & 0.666$^{***}$ & 0.641$^{***}$ & 1.049$^{***}$ & 0.917$^{***}$ \\ 
  & (0.099) & (0.107) & (0.098) & (0.099) & (0.096) & (0.099)  \\ 
 \% people with level 1 qualifications & 2.811$^{***}$ & 2.664$^{***}$ & 2.784$^{***}$ & 2.803$^{***}$ & 3.301$^{***}$ & 3.552$^{***}$ \\ 
  & (0.232) & (0.241) & (0.229) & (0.232) & (0.223) & (0.230) \\ 
 \% people with level 2 qualifications & $-$1.414$^{***}$ & $-$1.330$^{***}$ & $-$1.413$^{***}$ & $-$1.424$^{***}$ & $-$1.063$^{***}$ & $-$0.980$^{***}$ \\ 
  & (0.195) & (0.203) & (0.192) & (0.195) & (0.188) & (0.193) \\ 
 \% people with apprenticeships & $-$1.365$^{***}$ & 0.274 & $-$1.215$^{***}$ & $-$1.358$^{***}$ & $-$0.562$^{*}$ & $-$0.244 \\ 
  & (0.308) & (0.319) & (0.304) & (0.308) & (0.297) & (0.306) \\ 
 \% people with level 3 qualifications & $-$0.969$^{***}$ & $-$0.877$^{***}$ & $-$0.960$^{***}$ & $-$0.958$^{***}$ & $-$0.933$^{***}$ & $-$1.036$^{***}$ \\ 
  & (0.117) & (0.113) & (0.115) & (0.117) & (0.112) & (0.115) \\ 
 \% people providing 20+ hours unpaid care & $-$2.858$^{***}$ & $-$2.217$^{***}$ & $-$2.495$^{***}$ & $-$2.893$^{***}$ & $-$3.556$^{***}$ & $-$3.798$^{***}$ \\ 
  & (0.425) & (0.422) & (0.420) & (0.425) & (0.410) & (0.421) \\ 
 \% people in living in larger HHs & $-$0.720$^{***}$ & $-$0.068 & $-$0.687$^{***}$ & $-$0.724$^{***}$ & $-$0.826$^{***}$ & $-$0.820$^{***}$ \\ 
  & (0.106) & (0.103) & (0.104) & (0.106) & (0.102) & (0.103) \\ 
 Density of the LSOA & $-$0.001$^{***}$ & $-$0.0001 & $-$0.001$^{***}$ & $-$0.001$^{***}$ & $-$0.001$^{***}$ & $-$0.001$^{***}$ \\ 
  & (0.0001) & (0.0001) & (0.0001) & (0.0001) & (0.0001) & (0.0001) \\ 
  Presence of TT food bank &  &  & 50.633$^{***}$ &  & 41.552$^{***}$ & 38.849$^{***}$ \\ 
  &  &  & (1.902) &  & (1.869) & (1.893)\\ 
 Presence of IFAN food bank &  &  &  & $-$16.730$^{***}$ &  &  \\ 
  &  &  &  & (2.697) & \\ 
 Minimum distance to TT food bank &  &  &  &  &  $-$2.856$^{***}$ &  \\ 
  &  &  &  &  & (0.077) & \\ 
 Minimum public transport travel time &  &  &  &  &  & $-$0.206$^{***}$ \\ 
  &  &  &  &  &  & (0.005) \\ 
 Constant & $-$17.265$^{***}$ & $-$2.470 & $-$18.218$^{***}$ & $-$16.921$^{***}$ & $-$15.416$^{***}$ & $-$9.467$^{***}$  \\ 
  & (2.848) & (8.699) & (2.809) & (2.846) & (2.736) & (2.808) \\ 
\hline \\[-1.8ex] 
Fixed effects & No & Yes & No & No & No & No \\
R$^{2}$ & 0.182 & 0.340 & 0.204 & 0.183 & 0.245 & 0.252 \\ 
Adjusted R$^{2}$ & 0.182 & 0.335 & 0.204 & 0.183 & 0.245 & 0.252 \\ 
Residual Std. Error & 55.459 & 50.008 & 54.698 & 55.418 & 53.267 & 53.751 \\ 
DF & 25267 & 25057 & 25266 & 25266 & 25265 & 24324 \\
\hline \\[-1.8ex] 
\textit{Note:}  & \multicolumn{2}{r}{$^{*}$p$<$0.1; $^{**}$p$<$0.05; $^{***}$p$<$0.01} \\ 
\end{tabular} 
\end{table} 

\begin{table}[!t] \centering 
  \caption{Multiple regression models for predicting food bank use in rural areas with predictor variables and accessibility measures added} 
  \label{regression_rural_unstandardised} 
    \tiny
\begin{tabular}{@{\extracolsep{5pt}}lcccccc} 
 & \multicolumn{6}{c}{\textit{Dependent variable: Parcels per 1000 residents}} \\ 
\cline{2-7} 
\\[-1.8ex] & R1 & R2 & R3 & R4 & R5 & R6 \\ 
\hline \\[-1.8ex] 
\% of people who claim Universal Credit & 5.113$^{***}$ & 6.203$^{***}$ & 4.518$^{***}$ & 5.287$^{***}$ & 4.682$^{***}$ & 4.783$^{***}$ \\ 
  & (0.346) & (0.318) & (0.341) & (0.349) & (0.323) & (0.360) \\ 
 \% of single family HHs with a lone parent & 0.749$^{***}$ & 0.610$^{***}$ & 0.695$^{***}$ & 0.734$^{***}$ & 0.328 & 0.378 \\ 
  & (0.224) & (0.208) & (0.219) & (0.224) & (0.208) & (0.244) \\ 
 \% of people who are students & $-$0.042 & 0.505$^{***}$ & 0.026 & $-$0.028 & $-$0.231 & $-$0.149 \\ 
  & (0.167) & (0.155) & (0.164) & (0.167) & (0.156) & (0.186) \\ 
 \% of HHs that are socially rented & 0.626$^{***}$ & 0.506$^{***}$ & 0.620$^{***}$ & 0.627$^{***}$ & 0.656$^{***}$ & 0.698$^{***}$ \\ 
  & (0.059) & (0.059) & (0.058) & (0.059) & (0.055) & (0.065) \\ 
  \% of HHs with 1+ person with a disability & 2.343$^{***}$ & 1.919$^{***}$ & 2.088$^{***}$ & 2.358$^{***}$ & 2.260$^{***}$ & 2.801$^{***}$ \\ 
  & (0.226) & (0.217) & (0.222) & (0.226) & (0.211) & (0.247) \\ 
  \% people who are economically inactive & $-$0.647$^{***}$ & $-$0.396$^{***}$ & $-$0.648$^{***}$ & $-$0.652$^{***}$ & $-$0.828$^{***}$ & $-$1.034$^{***}$ \\ 
  & (0.119) & (0.111) & (0.116) & (0.119) & (0.111) & (0.129) \\ 
 \% people who work part time & 0.762$^{***}$ & $-$0.176 & 0.841$^{***}$ & 0.767$^{***}$ & 1.636$^{***}$ & 1.924$^{***}$ \\ 
  & (0.130) & (0.147) & (0.128) & (0.130) & (0.123) & (0.147) \\ 
  \% people with level 1 qualifications & 0.145 & 1.237$^{***}$ & 0.282 & 0.134 & 0.391 & 0.905$^{***}$ \\ 
  & (0.295) & (0.313) & (0.290) & (0.295) & (0.275) & (0.323) \\ 
  \% people with level 2 qualifications & 0.553$^{**}$ & $-$0.069 & 0.558$^{**}$ & 0.540$^{**}$ & 0.947$^{***}$ & 1.080$^{***}$ \\ 
  & (0.268) & (0.254) & (0.262) & (0.267) & (0.249) & (0.294) \\ 
  \% people with apprenticeships & $-$0.920$^{**}$ & 0.078 & $-$0.506 & $-$0.919$^{**}$ & $-$0.101 & $-$1.133$^{***}$ \\ 
  & (0.387) & (0.385) & (0.380) & (0.387) & (0.360) & (0.426) \\ 
  \% people with level 3 qualifications & $-$0.251 & $-$0.782$^{***}$ & $-$0.354$^{**}$ & $-$0.263 & $-$0.411$^{**}$ & $-$0.802$^{***}$ \\ 
  & (0.183) & (0.170) & (0.179) & (0.183) & (0.170) & (0.198) \\ 
  \% people providing 20+ hours  unpaid care & $-$3.482$^{***}$ & $-$1.790$^{***}$ & $-$3.005$^{***}$ & $-$3.522$^{***}$ & $-$3.636$^{***}$ & $-$5.417$^{***}$ \\ 
  & (0.538) & (0.519) & (0.528) & (0.538) & (0.502) & (0.592) \\ 
  \% people in living in larger HHs & $-$0.015 & 0.251 & 0.055 & $-$0.037 & 0.064 & 0.093 \\ 
  & (0.201) & (0.195) & (0.198) & (0.201) & (0.187) & (0.210) \\ 
  Density of the LSOA & 0.0005$^{**}$ & 0.001$^{***}$ & 0.0005$^{**}$ & 0.0005$^{**}$ & $-$0.0002 & $-$0.001$^{***}$ \\ 
  & (0.0002) & (0.0002) & (0.0002) & (0.0002) & (0.0002) & (0.0002) \\ 
  Presence of TT food bank &  &  & 49.902$^{***}$ & & 37.847$^{***}$ & 29.343$^{***}$ \\ 
  &  &  & (2.328) &  & (2.235) & (2.422) \\ 
  Presence of IFAN food bank &  &  &  & $-$12.989$^{***}$ &  &   \\ 
  &  &  &  & (3.512) &  & \\ 
  Minimum distance to TT food bank &  &  &  &  & $-$1.852$^{***}$ & \\ 
  &  &  &  &  & (0.052) &  \\ 
  Minimum public transport travel time &  &  &  &  &  & $-$0.194$^{***}$ \\ 
  &  &  &  &  &  & (0.006) \\ 
  Constant & $-$23.706$^{***}$ & 1.027 & $-$27.594$^{***}$ & $-$23.308$^{***}$ & $-$28.218$^{***}$ & $-$15.428$^{**}$ \\ 
  & (5.791) & (8.342) & (5.681) & (5.789) & (5.390) & (6.280) \\ 

\hline \\[-1.8ex] 
Fixed effects & No & Yes & No & No & No & No \\
R$^{2}$ & 0.169 & 0.374 & 0.201 & 0.170 & 0.281 & 0.308 \\ 
Adjusted R$^{2}$ & 0.168 & 0.366 & 0.200 & 0.168 & 0.280 & 0.307 \\ 
Residual Std. Error & 48.974 & 42.740 & 48.019 & 48.946 & 45.559 & 47.213 \\ 
DF & 11408 & 11285 & 11407 & 11407 & 11406 & 8871 \\ 
\hline \\[-1.8ex] 
\textit{Note:}  & \multicolumn{2}{r}{$^{*}$p$<$0.1; $^{**}$p$<$0.05; $^{***}$p$<$0.01} \\ 
\end{tabular} 
\end{table}

\end{document}